\def\year{2020}\relax
\documentclass[letterpaper]{article} %DO NOT CHANGE THIS
\usepackage{aaai20}  %Required
\usepackage{times}  %Required
\usepackage{helvet}  %Required
\usepackage{courier}  %Required
\usepackage{url}  %Required
\usepackage{graphicx}  %Required
\usepackage{graphicx}  %Required

\usepackage{booktabs}
\usepackage{multirow}
\usepackage{array}
\usepackage{siunitx}
\usepackage{subfigure}
\usepackage{amsmath}
\usepackage{supertabular} 
\usepackage{booktabs}
\usepackage[table]{xcolor}
\usepackage{tablefootnote}
\usepackage{xspace}
\usepackage{dblfloatfix}
\newcolumntype{L}[1]{>{\raggedright\let\newline\\\arraybackslash\hspace{0pt}}m{#1}}
\newcolumntype{C}[1]{>{\centering\let\newline\\\arraybackslash\hspace{0pt}}m{#1}}
\newcolumntype{R}[1]{>{\raggedleft\let\newline\\\arraybackslash\hspace{0pt}}m{#1}}

\newcounter{RQCounter}

\newcommand{\RQ}[2]{%
\refstepcounter{RQCounter} \label{#1}
    \noindent\textbf{RQ\arabic{RQCounter}.#2}
}

\newcommand*\samethanks[1][\value{footnote}]{\footnotemark[#1]}

\title{Driving the Last Mile: Characterizing and Understanding Distracted Driving Posts on Social Networks}

\author{
Hemank Lamba$^1$, Shashank Srikanth$^{2}\thanks{Equal authors by contribution}$, Dheeraj Reddy Pailla $^{2}\samethanks$, Shwetanshu Singh$^3$,	\\
\Large\bf{Karandeep Juneja}$^2$ Ponnurangam Kumaraguru$^3$	\\
$^1$Carnegie Mellon University, USA, 
$^2$International Institute of Information Technology, Hyderabad, India \\
hlamba@cs.cmu.edu, \{shashank.s, dheerajreddy.p, karandeepsingh.juneja\}@students.iiit.ac.in \\
$^3$Indraprastha Institude of Information Technology, Delhi, India	\\
\{shwetanshus, pk\}@iiitd.ac.in
}

\begin{document}
%\urlstyle{rm}

\newcommand{\driving}{distracted driving content }
\newcommand{\Driving}{Distracted driving content }
\newcommand{\DRIVING}{\textsc{Distracted Driving Content}\xspace}

\newtheorem{observation}{Insight}
\newtheorem{hyp}{Hypothesis}

\maketitle
\begin{abstract}
In 2015, $391$,$000$ people were injured due to distracted driving in the US. One of the major reasons behind distracted driving is the use of cell-phones, accounting for $14\%$ of fatal crashes. Social media applications have enabled users to stay connected, however, the use of such applications while driving could have serious repercussions - often leading the user to be distracted from the road and ending up in an accident. In the context of impression management, it has been discovered that individuals often take a risk (such as teens smoking cigarettes, indulging in narcotics, and participating in unsafe sex) to improve their social standing. Therefore, viewing the phenomena of posting distracted driving posts under the lens of self-presentation, it can be hypothesized that users often indulge in risk-taking behavior on social media to improve their impression among their peers. In this paper, we first try to understand the severity of such social-media-based distractions by analyzing the content posted on a popular social media site where the user is driving and is also simultaneously creating content. To this end, we build a deep learning classifier to identify publicly posted content on social media that involves the user driving. Furthermore, a framework proposed to understand factors behind voluntary risk-taking activity observes that younger individuals are more willing to perform such activities, and men (as opposed to women) are more inclined to take risks. Grounding our observations in this framework, we test these hypotheses on $173$ cities across the world. We conduct spatial and temporal analysis on a city-level and understand how \driving posting behavior changes due to varied demographics. We discover that the factors put forth by the framework are significant in estimating the extent of such behavior.
%We test these hypotheses on our dataset by analyzing data collected from $130$ cities all over the world. We find that there exists an inverse correlation with the percentage of youth population and the mean income. Further, there is a positive correlation with the gender ratio. \sshashank{The last two sentences seem abrupt}
\end{abstract}

\section{Introduction}
Distracted driving is any non-driving activity that the driver engages in, which can lead to visual (taking eyes off the road), manual (taking hands off the driving wheel) or cognitive (taking the mind off driving) distractions~\cite{definition2017distracted}. Distracted driving is particularly risky: In 2015, fatal crashes involving distracted drivers resulted in the deaths of $9$ individuals and $1$,$000$ injuries in the US alone~\cite{national2017distracted}. 

Usage of cell-phones while driving has been a primary reason for distraction-affected crashes, resulting in 69,000 total crashes in 2015~\cite{national2017distracted}. Texting while driving can be particularly devastating as it combines all three types of distractions (visual, manual, and cognitive)~\cite{vegega2013understanding,lipovac2017mobile,caird2008meta,horrey2006examining}. Among cell-phone users, teenagers and young adults are especially at risk. Studies show that $42$\% of high schoolers text multiple times while driving~\cite{kann2016youth}, and teenagers and young adults comprise $36$\% of distracted drivers using cell phones~\cite{national2017distracted}. 

We argue that social media use can have similar effects. 
Individuals spend $30$\% of their weekly online time on social networking
applications~\cite{gwi}, with $78$\% of traffic coming from smartphones%
~\cite{nielsen}. For instance, an average Snapchat user spends $30$ minutes daily on the platform~\cite{snapchat18}. However, while many studies investigated the risk of using cell phones while driving, prior work generally focused on texting and emailing; thus, the impact of social media use remains relatively unknown. 

We address this gap in our paper, by using large-scale data from Snapchat to develop a deep-learning based classifier to classify a  post as \driving or not. Then, grounded in Lyng's edgework theory~\cite{edgework1990lyng} (details in the next section), we investigate the extent to which people create and post content while driving 
%\hl{Moving this to somewhere later in the draft as first page gets shared a lot} or while being in the passenger seat, \footnote{Arguably a front-seat passenger creating social media content, e.g., a video, could also be a source of distraction for the driver.}
and characterize the users and spatial and temporal patterns associated with
higher incidence of such content.

% In this paper we address this gap. 
% Using large-scale data from Snapchat, a popular social media platform, 
% we develop a deep-learning-based classifier to detect whether a post can 
% be classified as \driving or not.

%\hl{We find that ...}
We discover that (1) a deep learning classifier trained on content has a good performance in detecting distracted driving content, (2) distracted driving content posting behavior is widespread - 23\% of snaps posted are related to distracted driving. Further, by analyzing the spatial and temporal patterns, we discovered that (3) distracted driving content is generally posted in night-time and regional affects are visible in the temporal patterns of such behavior and (4) distracted driving content posts are concentrated to only certain spots in the city. Finally, we also discovered that age and gender play a key role in inferring who is more likely to participate in such risk-taking behavior.

In summary, we make the following main contributions:
(1)~a classifier to detect \driving posting behavior on Snapchat;
(2)~an empirical study characterizing the extent of \driving behavior across $173$ cities around the world, the types of users more likely to engage in such behavior, and spatial and temporal patterns of \driving snaps in these cities. %To the best of our knowledge, this is the first work that tackles the concept of voluntary risk-taking behavior on social media platforms.

Our results have implications for platform designers and policymakers. Our proposed deep-learning based classifier can identify \driving content posted on social media. Furthermore, the spatial and temporal patterns and individual user characteristics we uncover can inform the design of region-specific interventions for certain cities where such behavior is common, and for specific times when users generate these posts; as well as the design of individual-level interventions and educational campaigns for at-risk populations.

\textbf{Privacy and Ethics}:
We collect data from SnapMaps, a geographical interface for Snapchat, which is publicly available. The data posted on the platform is already anonymized, and we neither collect nor use any personally identifiable information for our analysis. For variables extracted from the census, we only use the variables as is collected by the respective country's census department.
%\hl{REFINE}No privacy laws have been violated; everything is public and open-sourced.

\textbf{Data and Code}:
Our code and data is publicly available at \url{http://precog.iiitd.edu.in/research/distracted_driving/}.

\section{Development of Research Questions}

Our work is grounded in two theoretical frameworks. First, Goffman's dramaturgical theory~\cite{goffman1959presentation} describes how individuals may engage in risk-taking behavior to improve their peers' impressions of them, even when interacting through online social media platforms~\cite{hogan2010presentation}. Goffman introduced the term ``impression management", which has been widely used to explain how an individual presents an idealized rather than a more authentic version of themselves~\cite{goffman1959presentation}. In the context of risk-taking behavior, Leary et al.~\cite{leary1994self} analyzed voluntary risk-taking activities such as avoiding condoms, indulging in narcotics and steroid use, and reckless driving, and suggested that such risk-taking activities are undertaken to improve the impression of individuals among their peers~\cite{leary1994self}. Hogan~\cite{hogan2010presentation} extended Goffman's concept of impression management to online social media websites and considered the online social media platforms as a stage that allows users to control their impressions via status messages, pictures posted, and social media profiles. Similarly, we expect that social media users could post distracted driving content. Therefore, we ask our first research question:

\noindent
\mbox{\RQ{rq1}{}}\textbf{[Extent]} \emph{What is the extent of 
distracted driving content posting behavior on Snapchat?}

\smallskip
Second, Lyng's edgework theory~\cite{edgework1990lyng} characterizes voluntary risk-taking behavior (or, edgework) and identifies a range of individual and social factors that characterize the edgeworkers. The framework defines edgework activities as those where there is a ``clearly observable threat to one's physical or mental well-being",  such as rock-climbing, auto-racing, criminal behavior, drug use, etc. 
Edgework theory is social psychological, resting on the idea that 
individuals indulge in such activity to maintain the ``illusion of 
control."  Treating illusory sense of control as a factor, Lyng observed that edgework is more common among young people than among older people and among males than females. Other studies have found similar evidence related to the gender and age of the risk-takers~\cite{doyle1995male,leary1994self}. Building on this line of work, we also investigate if the demographic factors put forward by edgework framework also hold for \driving posting behavior on Snapchat. We therefore ask:

\noindent
\mbox{\RQ{rq2}{}}\textbf{[Demographics]} \emph{Which user demographic characteristics correlate with posting distracted driving content?}

\smallskip
Besides the individual characteristics, Lyng also noted that individuals who are under pressure from external social forces are also more inclined to do edgework, as a way to exhibit control over experiences that are potentially even more dangerous. We expect that different geographic locations can give indications about the culture in that particular part of the world and hence the social forces at play. In addition, social media use is known to vary across geographies~\cite{hochman2012visualizing,kim2011cultural,tifentale2015selfiecity}. For example, Kim et al.~\cite{kim2011cultural} studied how cultural contexts influence usage of social network sites among teenagers from US and Korea, finding that Korean participants used it for receiving acceptance from their peers, while US participants used the websites only for entertainment purposes. Similar studies were carried out by Hochman et al.~\cite{hochman2012visualizing} and Tifentale et al.~\cite{tifentale2015selfiecity}, where they noticed different patterns across geographies in terms of photo-sharing behavior. A better understanding of the geographic patterns can help in designing more appropriate and effective interventions for the at-risk population in such regions. We, thus ask:

\noindent
\mbox{\RQ{rq3}{}}\textbf{[Spatial Analysis]} \emph{How does distracted driving content posting behavior vary across cities worldwide?}

\smallskip
% Another common dimension of variability in social media use is temporal.
There is much variability in the temporal patterns of social media usage. For example, Golder et al.~\cite{golder2007rhythms} analyzed Facebook messaging pattern across universities and discovered temporal rhythms. They showed that students across all universities followed a ``weekday" and a ``weekend" pattern and further showed that students in the same university behaved similarly. Grinberg et al.~\cite{grinberg2013extracting} discovered interpretable temporal patterns for mention of different terms related to nightlife, coffee, etc.\ on Twitter and Foursquare checkins. Golder et al.~\cite{golder2011diurnal} further analyzed the temporal patterns of Twitter messages and were able to identify diurnal and seasonal mood rhythms, such as observing that people were generally happier on weekends; and that the morning peak in the number of messages was delayed by 2 hours on weekends. We investigate whether we can derive similar diurnal patterns for \driving posting behavior, and ask:

\noindent
\mbox{\RQ{rq4}{}}\textbf{[Temporal Analysis]} \emph{How does distracted driving content posting behavior vary with time?}

\smallskip
However, before we can begin to study \driving posting behavior on
Snapchat empirically, we first need to be able to detect such behavior. A major component of our work building a classifier to identify \driving, where the content creator is driving or is distracted while driving. A popular stream of work in the area of classifying videos is to apply multiple image-based classifiers on the frames of the given video. To this end, He et al.~\cite{He_2016} proposed a deep learning model that learns the residual functions and out-perform previous competitors in a widely popular ImageNet challenge. Zagoruyko et al.~\cite{zagoruyko2016wide} further improved the ResNet model and proposed a Wide Residual Network (WRN), which uses the increased width of the network to improve accuracy. Xie et al.~\cite{Xie_2017} modified the ResNet model by introducing a new hyper-parameter called cardinality to better tune the depth and width of the model. We use some of these architectures as candidate models for our deep learning classifiers. Among the video classification approaches used for action recognition, an approach that operates on spatio-temporal 3D CNNs stands out~\cite{hara2018can} by having high accuracy on standard action recognition datasets such as Kinetics and UCF101. Based on the above insights, we explore the feasibility of learning a robust classifier to distinguish between \driving and non-distracted driving content, asking:

\noindent
\mbox{\RQ{rq0}{}}\textbf{[Detection]} \emph{How can we use Snapchat content to distinguish between distracted driving and other videos? Moreover, how accurate is such a classifier?}

\section{Data Collection and Dataset}
In this work, we study a widely used social media platform, Snapchat. Snapchat is a popular platform that allows users to post multimedia content(\emph{snaps}) that can be shared with other users - visible by all or only by friends. Our dataset is based on SnapMap - a unique feature where any content can be posted publicly anonymously. The content posted on Snap Map is automatically geo-tagged and is shown in a localized region, though not giving the exact location.

%Online social media platforms allow users to post, view, like, and comment on content posted by others. With the growth of social networking culture, multiple such platforms exist - e.g., Facebook, Instagram. However, most of the platforms do not provide easy access to such content for research purposes. In this paper, we make use of a popular social networking platform - Snapchat.

%Snapchat is a novel platform, which allows users to post multimedia content (\emph{snaps}), which can be shared with other users - publicly visible by all or by friends only. The novelty of the platform is that the content posted on it is ephemeral i.e., it can be viewed only for 24 hours from posting. 

%Snapchat is a novel and new platform for studying this phenomenon, given its philosophy to make the messages (snaps) inaccessible beyond a 24-hour window after posting. However, the recent release of Snap Maps (a global public feed of Snaps) ~\url{https://map.snapchat.com} opens up the possibility to better study the platform and characterize the behavior of its user group.

\begin{table}[t]
    \centering
    \caption{A sub-sample of the cities selected for analysis.}
    \begin{tabular}{lllll}
    \toprule
        City & \parbox[t]{0.8cm}{Economic\\Status\tablefootnote{\url{https://www.cia.gov/library/publications/the-world-factbook/appendix/appendix-b.html}}} & Pop. & \parbox[t]{0.8cm}{Male\\ (\% age)}  & \parbox[t]{0.8cm}{Pop.($<20$)} \\  
        \midrule
         Cape Town & Developed & $4.43$M & $48.90$ & $0.329$ \\
         London & Developed & $9.05$M & $49.80$ & $0.247$ \\
         Melbourne & Developed & $4.77$M & $49.00$ & $0.241$ \\
         New York & Developed & $8.58$M & $47.70$ & $0.232$ \\
         Rio De & Developing & $13.29$M & $46.80$ & $0.267$ \\
         Janeiro & & & & \\
         Riyadh & Developing & $6.91$M & $59.17$ & $0.220$ \\
    \bottomrule
    \end{tabular}
    \label{table:cities}
\end{table}

\subsection{Data Collection}
For obtaining the data through SnapMap, we leverage the underlying API to collect data across $173$ cities. We select these cities such that they give us a wide coverage over the entire world and they were constrained on having a minimum population of $200k$ each. Further, we filter out cities where there is limited or restricted Snapchat usage (for example, Chinese metropolises). A sampled list of some of the cities selected for this analysis, with certain attributes (that we use for future analysis) is provided in Table~\ref{table:cities}.\footnote{A full list of cities is available in the Supplementary, to be uploaded on acceptance.} We utilize the shapefiles obtained from OpenStreetMap\footnote{\url{https://www.openstreetmap.org}} to precisely define the region enclosed by a city. In the absence of a city's shapefile, we use its bounding box values instead.

% To collect the data from each city, we define a rectangular bounding box over 354 the entire city by considering it's south-west and north-east geographic coordinates obtained by Google's Geo-coding.  API.\footnote{\url{https://developers.google.com/maps/documentation/geocoding/}} 
This overall city's region/bounding box is divided into smaller tiles using a grid such that each tile is $1km \times 1km$. A similar approach has been previously used in geographical studies on Snapchat~\cite{juhaszanalyzing} which utilizes a tile of size $2.4km \times 2.4$km respectively. We periodically collect snaps posted in each of these grid tiles, crawling each city once every 8 hours. The data collected lists the time at which the snap was posted in Coordinated Universal Time (UTC), which we then convert to the local time-zone of the corresponding city to allow for uniformity in the temporal analysis.

\begin{table}[!h]
    \centering
    \caption{Brief description of the data collected.}
    \begin{tabular}{lc}
    \toprule
        Number of Snaps collected & 6,431,553 \\ 
        Number of cities scraped & $173$ \\ 
        \midrule
        Time of first Snap & 16-03-2019 00:00:00 \\ 
        Time of last Snap & 15-04-2019 23:38:57 \\  
        \midrule
        Most active city & Riyadh ($1$,$023$,$836$) \\  
        Least active city & Havana $(114)$ \\  
        \midrule
        Most active day & 13th April, 2019 ($288K$) \\  
        Least active day & 30th March, 2019 ($89K$) \\ 
        \% Snaps deleted & $2.98\%$ \\
        \midrule
    \bottomrule
    \end{tabular}
    \label{table:dataset}
\end{table}

Overall, a brief statistics of the collected dataset is given in Table~\ref{table:dataset}. We observed that $204$,$874$ snaps were deleted after posting and were not used in our analysis. Though our work is concentrated on Snapchat, it can be easily extended to most social media platforms where users post multimedia content (images/videos).

%We are achieving this by leveraging the underlying API for Snap Maps over $137$ cities globally. Cities with a projected population exceeding $x$ have been taken into consideration for data collection. The rectangular bounding box defined by the south-west and north-east geographic coordinates is collected for each city via Google's Geo-coding API. Further, this box is partitioned into a grid system of squares. To maintain uniformity square tiles of 2,500m x 2,500m are chosen. We partition the geography of the cities into a grid system and repeatedly sample snaps from each grid-tile for each city over the same period of time. The data gathered is then utilized to determine the extent to which the phenomenon is prevalent in a particular city. 

% \section{Detecting \driving}
\section{Detecting Distracted Driving Content}

\begin{table*}[!b]
\centering
\caption{Performance of various classification methods, using different base architectures on our ground truth dataset.}
\begin{tabular}{llcccc}
\toprule
 Type & Architecture & Accuracy & Precision & Recall & F1 Score  \\ 
 \midrule
 \multirow{3}{*}{\parbox[c]{2.5cm}{Image-Based \\ (Single Voting)}} 
 & ResNeXt-50 & $0.924 \pm 0.005$ & $0.780 \pm 0.015$ & $0.958 \pm 0.01$ & $0.859 \pm 0.007$ \\
 & ResNet-34 & $0.919 \pm 0.007$ & $0.774 \pm 0.02$ & $0.948 \pm 0.013$ & $0.851 \pm 0.009$  \\
  & WideResNet & $\mathbf{0.926 \pm 0.009}$ & $\mathbf{0.792 \pm 0.030}$ & $\mathbf{0.948 \pm 0.016}$ & $\mathbf{0.862 \pm 0.012}$ \\
 \midrule
 \multirow{3}{*}{\parbox[c]{2.5cm}{Image-Based \\ (Majority Voting)}} & ResNeXt-50 & $0.947 \pm 0.001$ & $0.902 \pm 0.008$ & $0.876 \pm 0.011$ & $0.888 \pm 0.004$ \\
  & ResNet-34 & $0.942 \pm 0.004$ & $0.896 \pm 0.02$ & $0.860 \pm 0.012$ &  $0.877 \pm 0.005$ \\  
  & WideResNet & $\mathbf{0.947 \pm 0.003}$ & $\mathbf{0.914 \pm 0.011}$ & $\mathbf{0.868 \pm 0.019}$ & $\mathbf{0.890 \pm 0.006}$  \\
 \midrule
 \multirow{3}{*}{Video-Based} 
  & ResNet-34 & $0.930 \pm 0.008$ & $0.860 \pm 0.044$ & $0.860 \pm 0.033$ & $0.857 \pm 0.013$ \\ 
  & ResNeXt-101 & $\mathbf{0.941 \pm 0.003}$ & $\mathbf{0.876 \pm 0.015}$ & $\mathbf{0.880 \pm 0.026}$ & $\mathbf{0.876 \pm 0.008}$  \\  
 \bottomrule
 %\cmidrule(l){3-11} 
\end{tabular}
\label{table:classifier}
\end{table*}

To be able to build a classification model, we need to have a ground truth dataset of snaps with labels marking each as either \driving or non-distracted driving content. We built an annotation portal (details of the portal provided in Supplementary), and asked annotators to provide labels for over $15$K snaps, randomly sampled from our dataset. We annotate each snap for \driving\footnote{This annotations might sometimes contain content shot by the passenger of the car. Arguably a front-seat passenger creating social media content, e.g., a video, could also be a source of distraction for the driver.} or non-distracted driving content and ensure that at least three annotators annotated each snap. We obtained a Fleiss-Kappa inter-annotator agreement rate of $0.85$, which signifies almost perfect agreement~\cite{fleiss1973equivalence}. A snap was assigned a ground-truth label of \driving if two or more annotators agree that it is a \driving snap. An anonymized example of \driving snap can be viewed at \url{https://rebrand.ly/driving-snap}. This snap is clearly dangerous as it is created by an individual who is driving and hence is classified as an example of distracted  driving.\footnote{Annotation portal screenshot in Supplementary}

\noindent
\textbf{Dataset.}
We randomly sample and split the manually annotated snaps into training and test set of $8$,$634$ (6,392 negative, 2,242 positive) and $1$,$479$ snaps (1,118 negative, 361 positive) respectively. We train our model using $5$ fold cross-validation on this so obtained dataset. The number of positive samples (distracted driving) in our training dataset is much less than the number of negative samples (non-distracted driving) which creates a class imbalance. 
%Similarly, our test set consists of 361 positive samples and 1118 negative samples.

%We train our model by performing $5$ fold cross-validation on the training set and use cross-validation for finding out the best hyper-parameters as well. The number of positive samples (driving) in our training dataset ($2242$) is much less than the number of negative samples (non-driving) ($6392$) which creates a class imbalance issue. The ratio of positive to negative samples in the testing dataset is around $30\%$, which is similar to the ratio in the training set. To further validate the performance of our trained classifiers, we test them on a held-out test consisting of \sshashank{$5,000$, @dheeraj confirm} samples. 

We experiment with two different kinds of classifiers - image-based and video-based. The main distinction between both types of approaches is that the image-based classifiers first converts the snap (a video) into frames, and then each frame is classified independently as either \driving or non-distracted driving content. Post classification of each frame, various aggregation techniques (single and majority voting) are used to obtain a single label for the entire snap. On the other hand, the video-based classification methods use the entire video as an input. 

% Suppose we use a rudimentary binary classifier that indiscriminately labels every video it comes across as one that does not involve any distracted driving content. By definition, this model would be $75.6\%$ accurate due to the aforementioned distribution of our test set. We establish this particular model as a baseline to compare our image-based and video-based classifiers against.

\noindent
\textbf{Image Based Methods.} 
We build our image-based methods over existing image-based deep learning architectures. We leverage the best-performing classifiers that have achieved high accuracy on ImageNet Large Scale Visual Recognition Challenge \cite{Russakovsky_2015}. The challenge consisted of 1.2M images covering $1$,$000$ classes. Specifically, we experiment with ResNet-34~\cite{He_2016} ($24.19\%$ top 1 error), ResNeXt-50~\cite{Xie_2017} ($22.2 \%$ top 1 error) and WideResNet-50 (WRN)~\cite{zagoruyko2016wide} ($21.9$\% top 1 error). The wide residual networks perform well as they decrease the depth of the network and increase its width to increase the representational power of the residual blocks. We pre-train these architectures on the ImageNet dataset, following which we fine-tune them on our annotated dataset using transfer learning. Such a technique is based on transfer learning and is efficient even when a small number of samples are used to fine-tune~\cite{zeiler2014visualizing}. The number of training samples in our dataset after converting the videos to frames is $69$,$125$, which is sufficient for transfer learning. To solve the class imbalance issue, we use data augmentation techniques such as random cropping and horizontal flipping to increase the number of driving frames shown to the network during training. For converting the snaps (videos) to frames, we sample a frame every second - every 30th frame per second (video's original playback rate is 30fps). For each frame, we obtain a label of whether it is \driving or non-distracted driving content. To obtain a single label for the entire snap, we use two aggregation techniques - (a) Majority voting and (b) Single voting. For majority voting, we classify the entire snap to be \driving if the majority of the frames are assigned to distracted driving class, whereas for single voting, we classify the entire snap as \driving content if we classify even a single frame as distracted driving content. We tune the hyper-parameters of these models using 5-fold cross-validation and report their accuracy, precision, recall and F1 score \footnote{We report the precision, recall and F1 score of the minor class in all our results} on the test set in Table~\ref{table:classifier}.

% \begin{figure}[!b]
% \vspace{-1em}
%     \centering
%     \includegraphics[width=\linewidth]{images/fig-robust-accuracy.pdf}
%     \vspace{-1em}
%     \caption{Effect of two different types of frame sampling methods, along with different threshold for label aggregation on accuracy of WideResNet architecture. Random Frame denotes random sampling of frames every second and Single Frame technique involves using the last (30th) frame for each second.}
%     \label{fig:robust-accuracy}
% \end{figure}

\begin{figure}[!h]
\centering
%\subfigure[]{\includegraphics[width=0.49\columnwidth]{images/random_frame_thresh.pdf}}
%\subfigure[]{\includegraphics[width=0.49\columnwidth]{images/single_frame_thresh.pdf}}
\subfigure[]{\includegraphics[width=0.49\columnwidth]{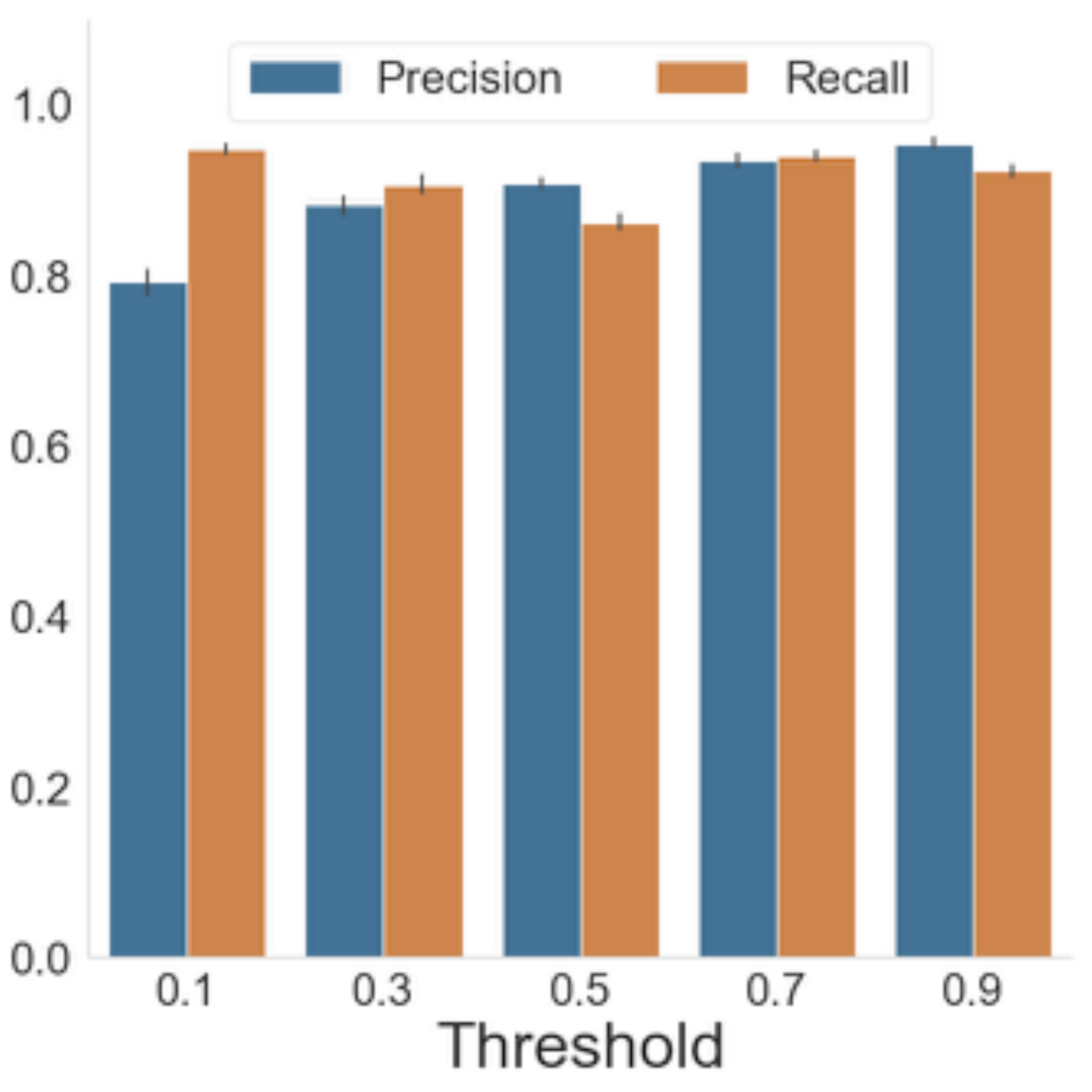}}
\subfigure[]{\includegraphics[width=0.49\columnwidth]{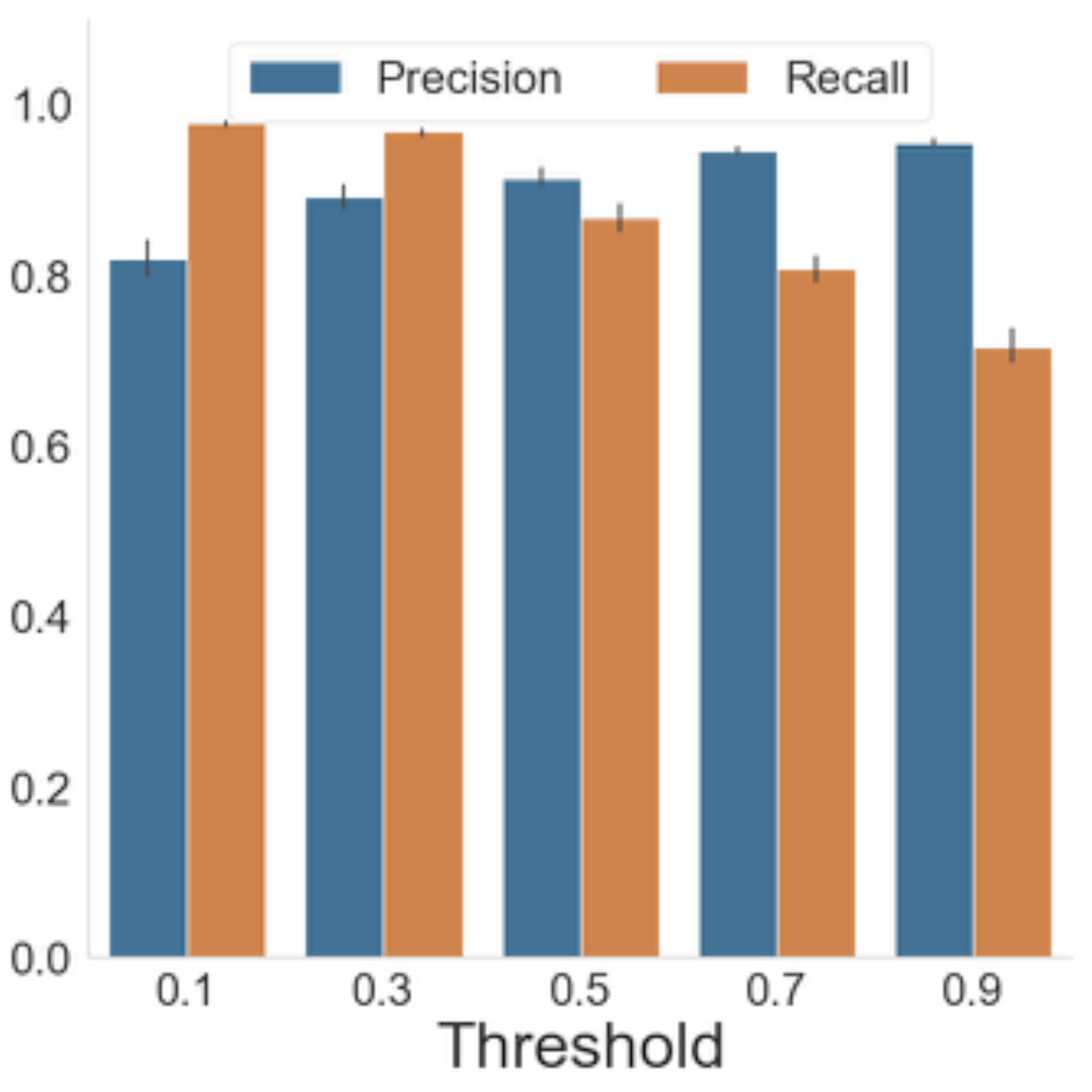}}
\vspace{-1em}
\caption{Precision and Recall for distracted driving class for (a) Random frame and (b) Single frame for different thresholds}
\label{fig:robust-accuracy}
\end{figure}

To measure the robustness of the frame selection, we compare our frame sampling strategy with that of a random frame sampling every second. We discover that the random frame sampling-based approach performs worse than our frame sampling strategy (random sampling has $93.8\%$, compared to our frame sampling's $94.8\%$ accuracy). Similarly, we also experiment with different voting aggregation techniques - where a snap is assigned a label if more than $10\%$, $30\%$, $50\%$, $70\%$ and $90\%$ of the frames have the same label. We report these results in Figure~\ref{fig:robust-accuracy}.

\noindent
\textbf{Video Based Methods.} For video-based classifiers, we again use state of the art architectures for a video classification task. Karpathy et al. explored multiple ways to fuse temporal information from consecutive frames using 2D pre-trained convolutions~\cite{karpathy2014large}. Similarly, Hara et al. proposed spatiotemporal 3D CNNs for video classification~\cite{hara2018can}. They examined deep architectures based on 3D Res-Net backbones for several datasets, achieving a top-5 accuracy of $85.7\%$ on the Kinetics dataset\cite{kay2017kinetics}. The Kinetics dataset consists of more than 300K videos with 400 class labels. To adapt these architectures for our classification task, we re-train two of their pre-trained models, which are based on ResNet34 and ResNeXt-101 architectures over our annotated dataset. Similar to the image-based methods, we utilize random cropping to solve the class imbalance issue. 

The image classifiers perform better than the video classifiers, as shown in Table~\ref{table:classifier}. We hypothesize that this might be because the image classifiers are pre-trained on the ImageNet dataset, which allows the classifiers to gain a much better internal representation of outdoor driving scenes. On the other hand, the Kinetics dataset on which the video classifier is pre-trained contains labels for action recognition tasks which do not transfer well to our task. Another reason why the video classifier does not perform as well as the image classifier is that the video classifiers require large amounts of data to train properly which, due to manual annotation limits, is not available for our dataset. 

\textbf{Training Details}
We train all our image-based models using Adam optimizer. The best model was trained with learning rate of $0.01$, batch size $16$, and utilized weight decay for regularization purposes. We train all the models for a maximum of $10$ epochs with the total training time of around $12$ hours on $4$ Nvidia GTX 1080Ti GPU. 

For the video classifier models, we use SGD (Stochastic Gradient Descent) with momentum and set the learning rate to $0.1$. We use a batch size of $32$ for the video classifiers and train both the models for a maximum of $60$ epochs each. We also use weight decay as a means of regularization for the model. 

\noindent
\textbf{Validation and Robustness of Classifier}
To validate the generalizability of our proposed method, we create a held-out test set from our collected dataset (dataset that was not previously used in any step of training). We randomly sampled $5$,$472$ snaps from our collected dataset ($1,404$ positive, $4,068$ negative). We did not place any geographic/temporal constraints on selecting these posts. On this held-out set, we see that all methods achieve a high accuracy of at least $0.93$, as shown in Table~\ref{table:holdout}.

\begin{table}[!h]
\centering
\caption{Performance of models on held-out set.}
\begin{tabular}{llll}
\toprule
 Type & Architecture & Accuracy & F1-Score  \\ 
 \midrule
 \multirow{3}{*}{\parbox[c]{2cm}{Image-Based \\ (Majority Voting)}} 
    & ResNeXt-50 & $\mathbf{0.953}$ & $\mathbf{0.91}$ \\
    & ResNet-34 & $0.948$ & $0.894$\\
    & WideResNet & $0.951$ & $0.904$ \\
 \midrule
 \multirow{2}{*}{Video-Based} 
    & ResNet-34 & $0.93$ & $0.859$\\ 
    & ResNeXt-101 & $\mathbf{0.942}$ & $\mathbf{0.859}$\\
 \bottomrule
\end{tabular}
\label{table:holdout}
\end{table}

In the above section, we show that our proposed deep learning approach that leverages the content of the snap can be used to detect \driving snaps successfully (RQ\ref{rq0}).

\section{Characterizing Temporal and Spatial Patterns}
In this section, we first measure the extent of \driving posting behavior across various cities on the platform. Temporal patterns have proven to be useful for analyzing trends; we perform temporal analysis on our dataset to understand when such type of behavior (posting distracted driving content) is prevalent. Further, we conduct spatial analysis to explore interesting patterns across and within each city to determine if such behavior is concentrated on certain parts of the city or is spread across uniformly.

\subsection{Extent of \driving}
\begin{figure}[]
%\vspace{-1em}
    \centering
    \includegraphics[width=0.95\linewidth]{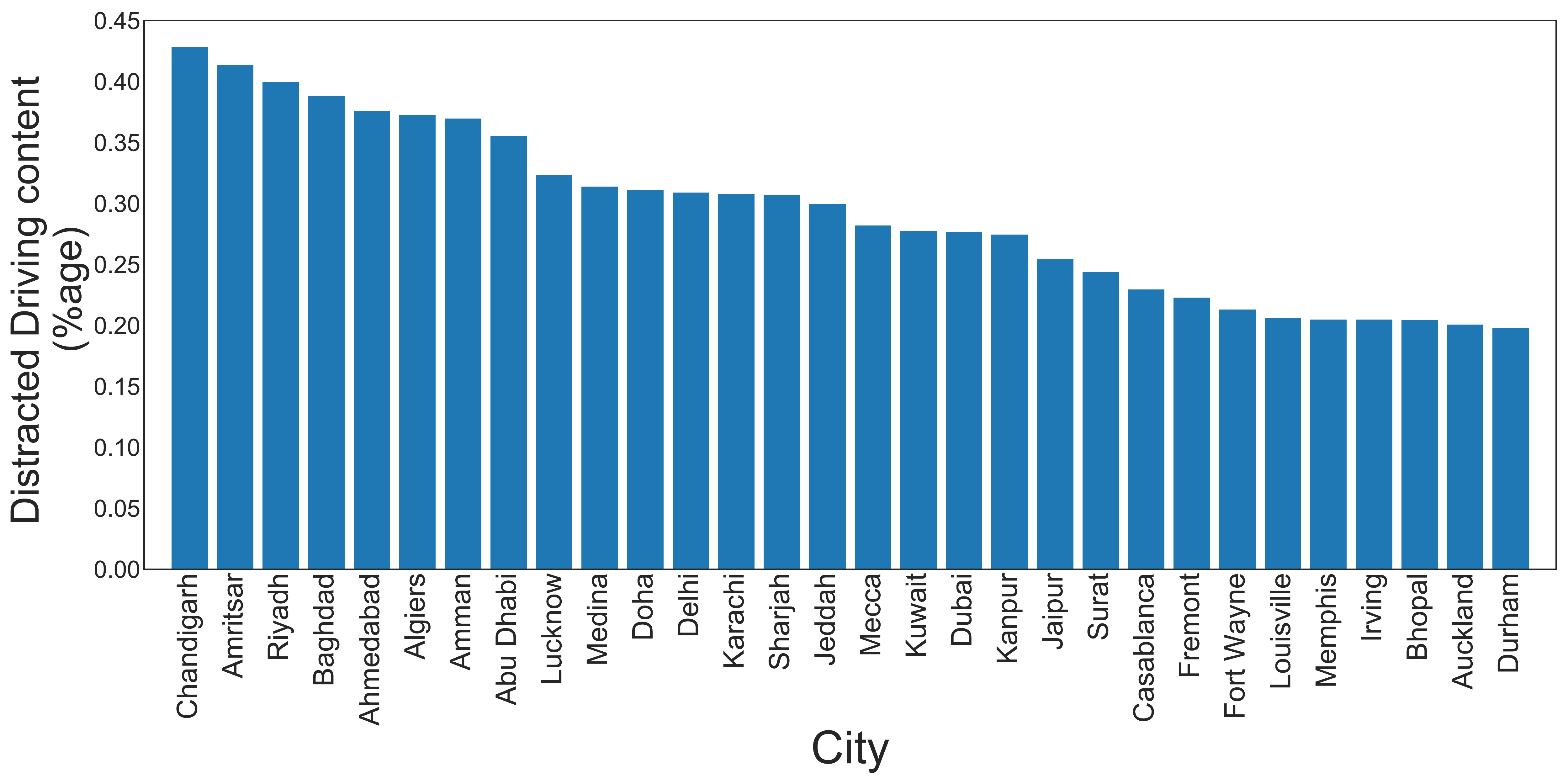}
%    \vspace{-3em}
    \caption{The top $30$ cities in our dataset ordered based on the ratio of driving snaps to the total snaps.}
    \label{fig:extent}
\end{figure}

Related to RQ\ref{rq1}, we want to understand the extent of posting \driving across various cities. To measure this, we applied our deep-learning classifier built in the previous section on all the snaps ($6.43$M) we collected. We discovered that around $23.56\%$ of the snaps in our dataset consisted of distracted driving content. Further, we analyzed which cities were exhibiting such behavior the most, and present it in Figure~\ref{fig:extent}. 

We observe that middle-eastern (Riyadh, Baghdad) and Indian cities (Chandigarh, Amritsar, Ahmedabad) were posting such content in high percentages ($>35\%$). Such behavior was found to be lower in European and American cities, and we find that there is not even a single European or American city in the list of top-$20$ cities. Moreover, the first American city (Fremont, CA) that has a high percentage ($22.26\%$) of distracted-driving content has only very few total numbers of snaps ($4$,$042$).

\begin{observation}[Regional Effect]
The trend of posting \driving on Snapchat is predominantly higher in Middle-Eastern and cities in Indian sub-continent, as compared to other cities across the world.
\end{observation}

%In this section, we characterize the spatial and temporal patterns emerging from the data, across multiple cities around the world. Temporal patterns prove to be useful to analyze trends in posting behavior and also to understand when such type of driving (risk-taking) behavior is prevalent. With spatial analysis, we understand the global patterns in posting behavior (both general and driving content) as well as explore within each city whether such behavior is concentrated or is uniform across the city limits. We applied the deep-learning classifier built in the earlier section on all the snaps we collected. We observed that the dataset comprised of $23.77\%$ driving-related snaps, which itself is a highly significant number.

\subsection{Temporal Analysis}
\begin{figure}[!t]
%\hfill
\centering
%\subfigure[]{\includegraphics[width=0.49\columnwidth]{images/temporal/diurnal_new.pdf}}
\subfigure[]{\includegraphics[width=0.49\columnwidth]{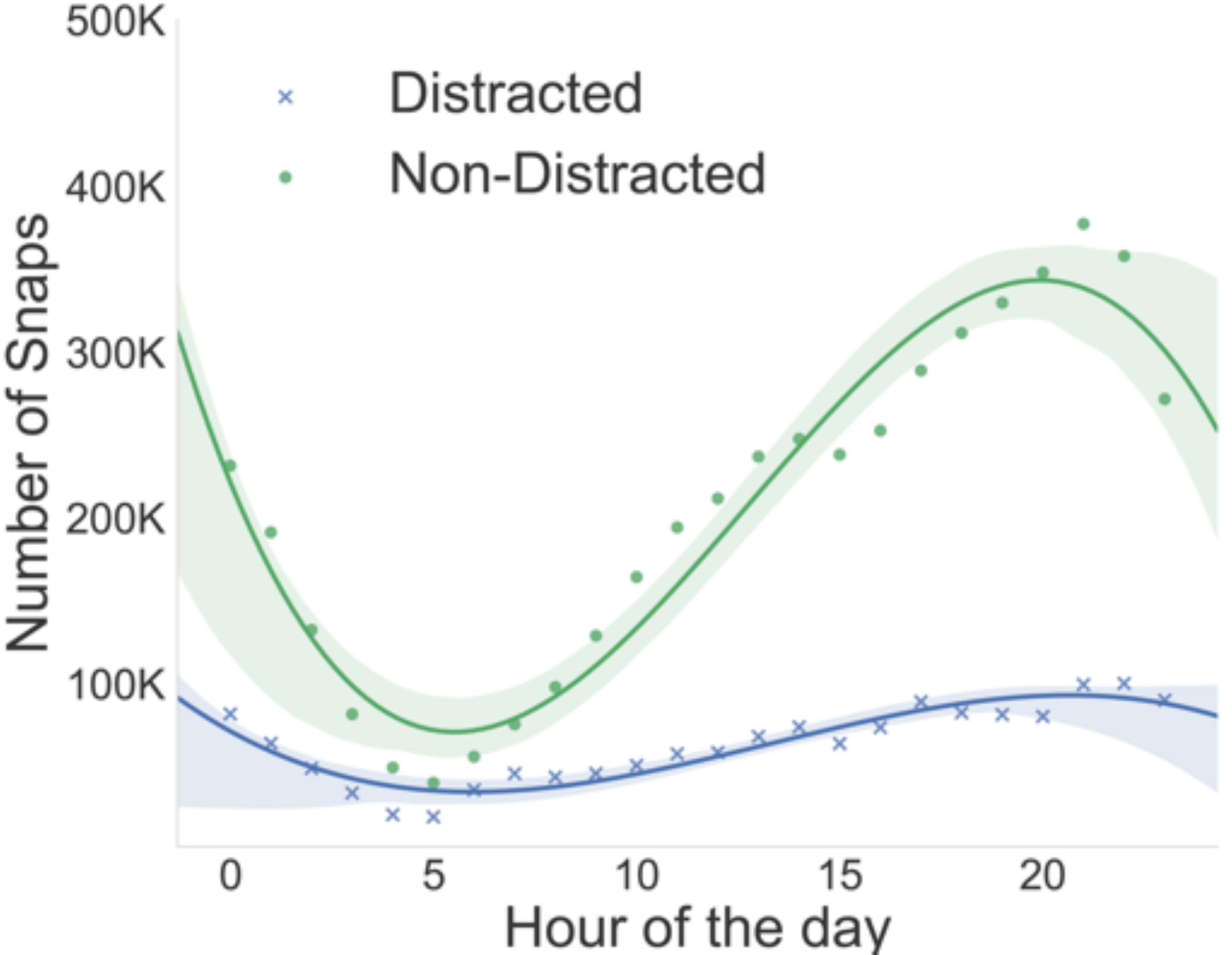}}
%\hfill
%\subfigure[]{\includegraphics[width=0.49\columnwidth]{images/temporal/Clustering_3_Hour_Day.pdf}}
\subfigure[]{\includegraphics[width=0.49\columnwidth]{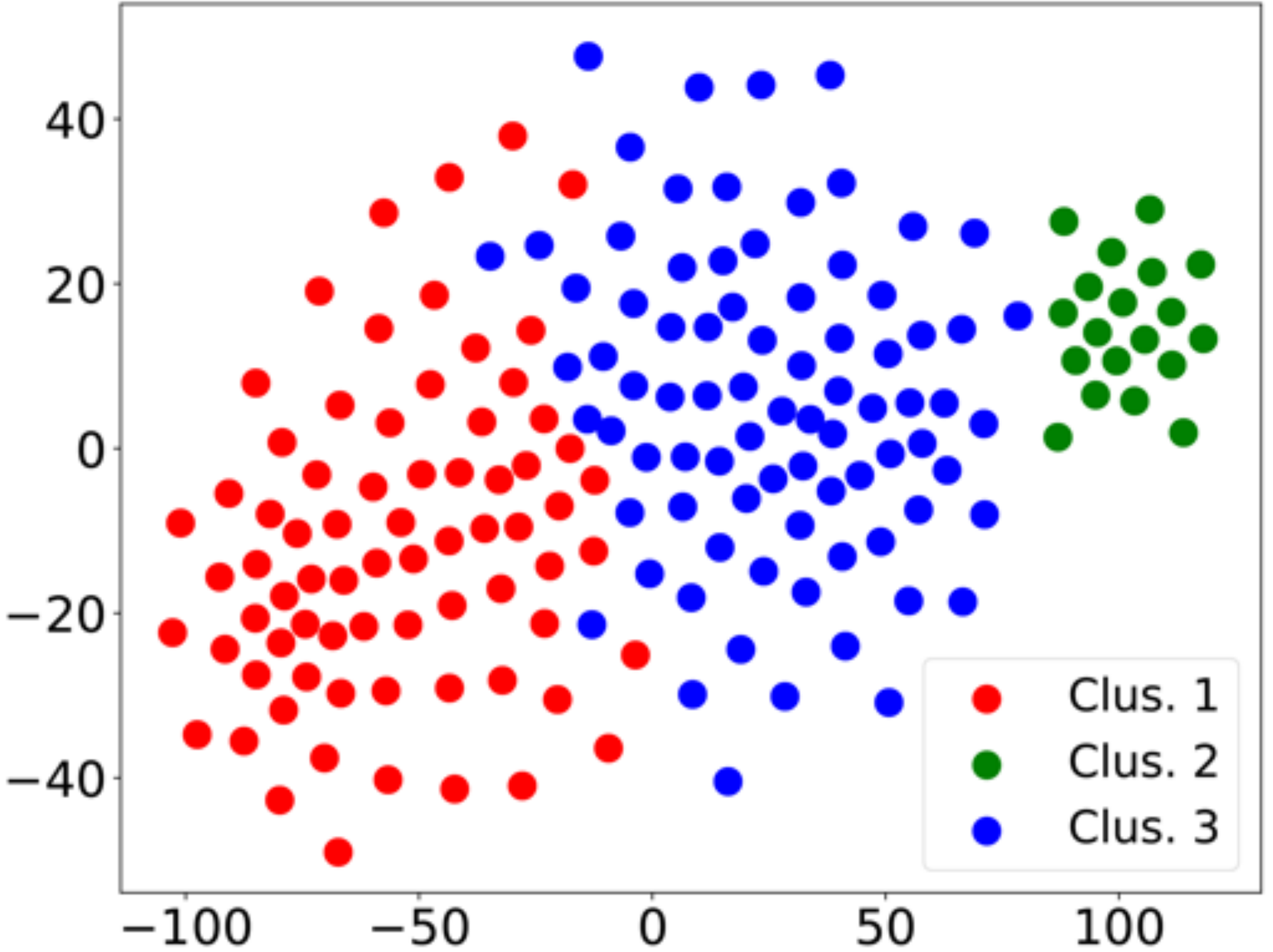}}
%\hfill
% \vspace{-1em}
\caption{(a) Diurnal trends (for both the distracted driving and non-distracted driving classes). The line plots denote the regression fit of the trends. (b) Cities clustered according to their temporal patterns.}
\label{fig:temporal_diurnal}
\end{figure}

We investigate how driving content posting behavior differs across time. In Figure~\ref{fig:temporal_diurnal}(a), we present the hour-wise distribution of (i) when users post distracted driving content, (ii) when users post any form of content. We can see that the \driving is approximately a uniform fraction of all the posts across the day. Users are often more active during the night-time (6PM-2AM), posting $73.51$\% more posts per hour in this period relative to the frequency of posting over other hours of the day. We observe a similar trend for driving snaps, where the number of driving snaps posted per hour during the evening to night window is found to be $77.83$\% more than the rest of the day. 

Further, to show that the driving snaps are a uniform fraction of the overall snaps, we compute the correlation between the number of driving snaps posted and the number of total snaps posted in every hour for the entire month of the data collected and find it highly correlated with a Pearson correlation coefficient of $0.9545$. We can also observe that a sharp drop in non-distracted-driving content is not complemented with a similar drop in the \driving posting. Due to this, we observe a pattern of higher distracted driving content posting activity through the night, and into the hours of the morning.

\begin{figure*}[t]
\centering
    %\hfill
%    \subfigure[Delhi]{\includegraphics[width=0.33\textwidth]{images/spatial/delhi-drive.pdf}}
    \subfigure[Delhi]{\includegraphics[width=0.33\textwidth]{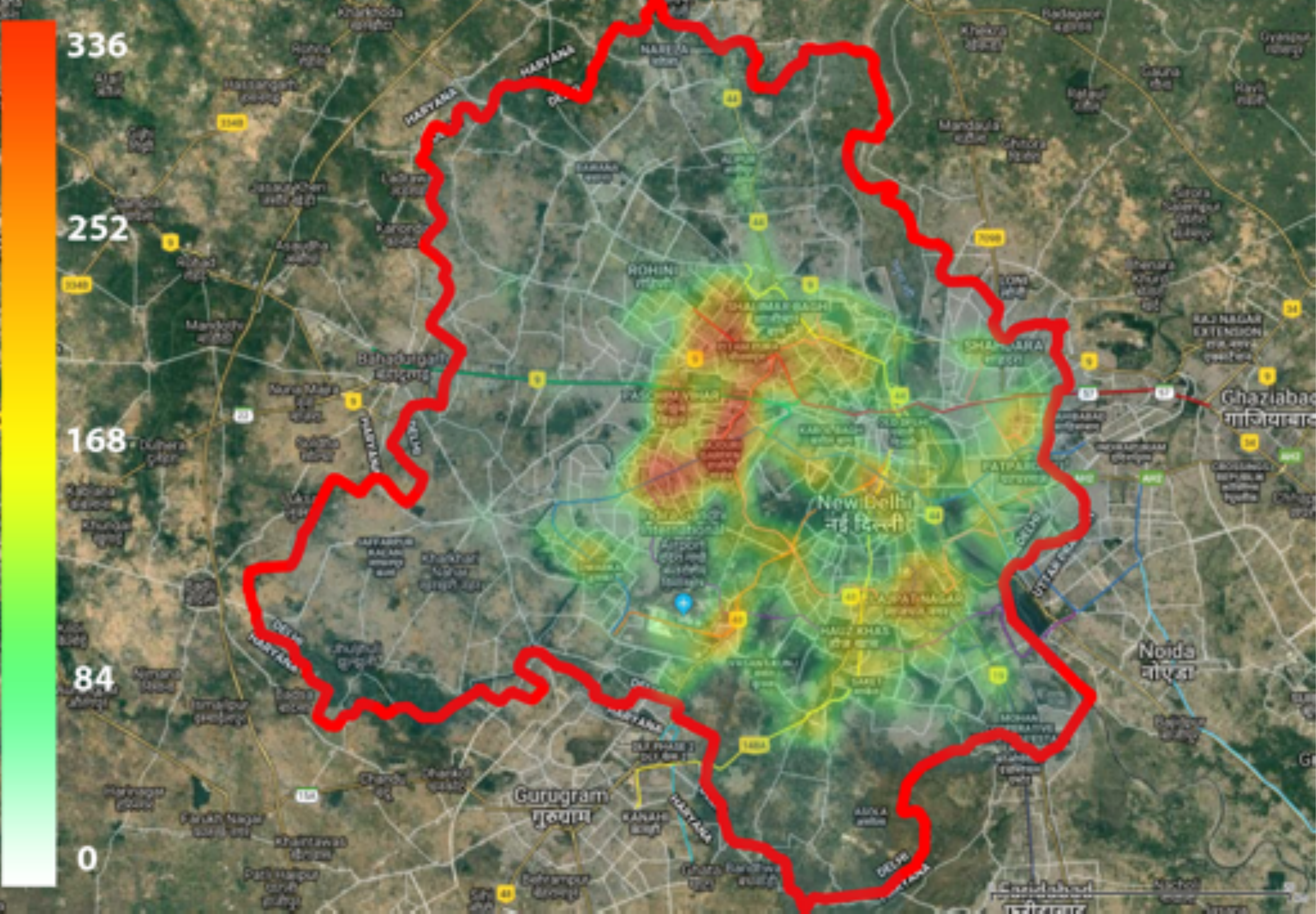}}
    %~\hfill
%    \subfigure[Riyadh]{\includegraphics[width=0.33\textwidth]{images/spatial/riyadh-drive.pdf}}
    \subfigure[Riyadh]{\includegraphics[width=0.33\textwidth]{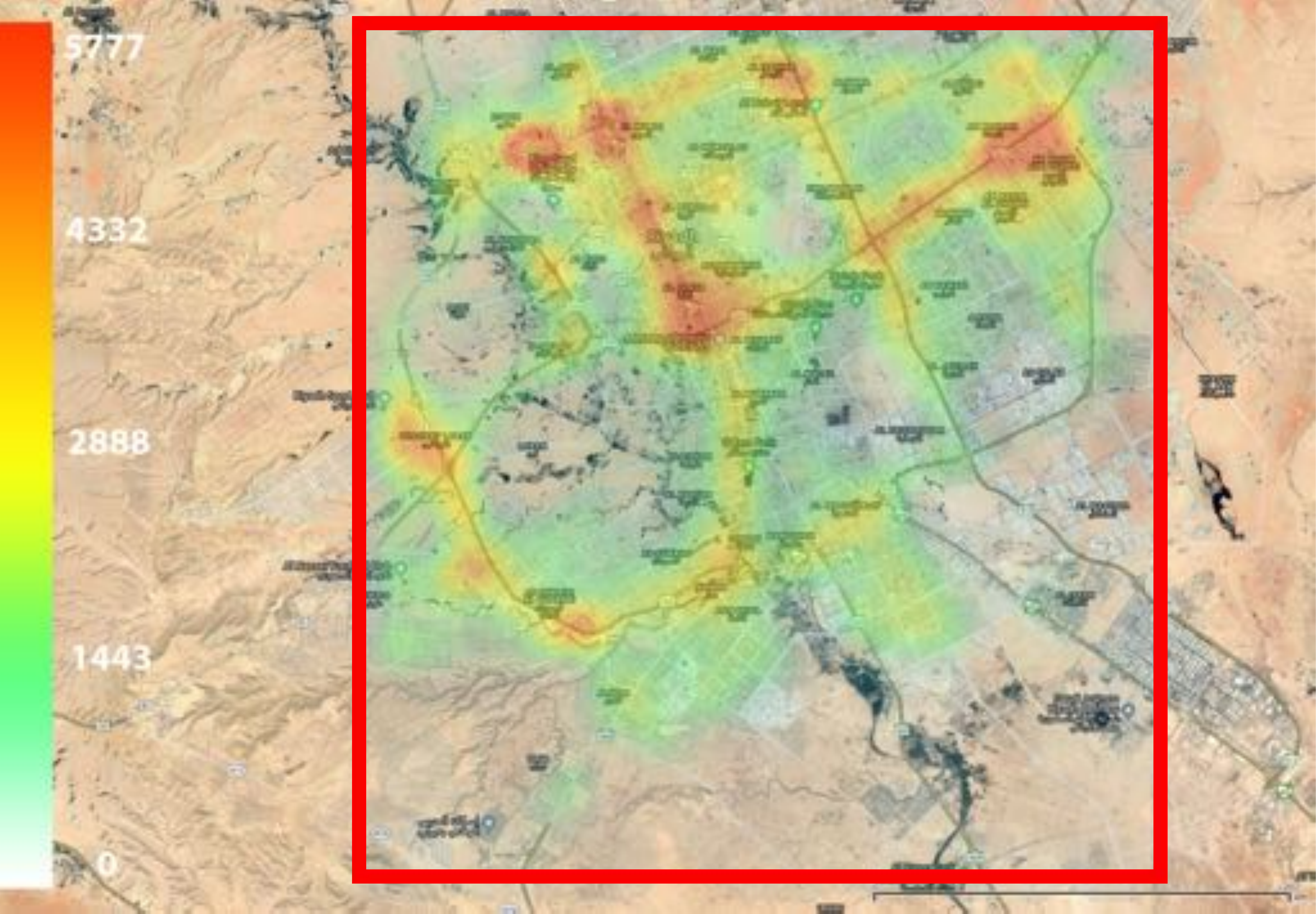}}
    %~\hfill
%    \subfigure[New York City]{\includegraphics[width=0.33\textwidth]{images/spatial/nyc-drive.pdf}}
    \subfigure[New York City]{\includegraphics[width=0.33\textwidth]{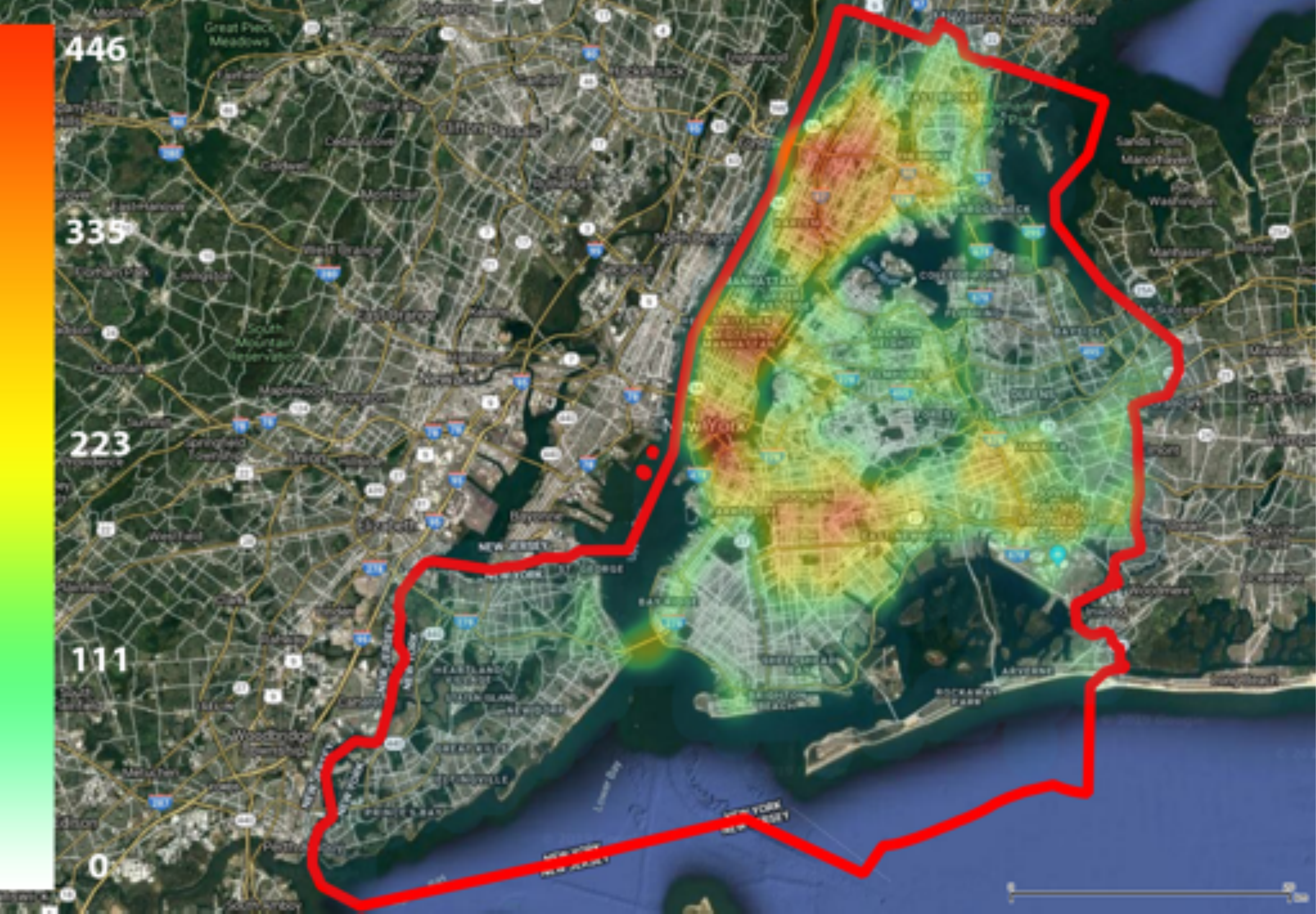}}
    \caption{Spatial analysis (frequency distribution plots) of three cities (from Table \ref{table:cities}. It can be noticed that the distracted driving behavior is concentrated around certain hotspots in the entire city.)}
    \label{fig:spatial_heatmaps}
\end{figure*}

\begin{observation}[Night-time Driving]
The incidence of posting while driving behavior over the night is more pronounced than other forms of content posting during the same hours.
\end{observation}

We further investigate the different temporal patterns that exist across different cities. We cluster the fraction of distracted driving snaps posted per hour over the entire week for each city. Using silhouette score coefficient~\cite{rousseeuw1987silhouettes} and also Elbow method~\cite{thorndike1953belongs}, we estimated the number of clusters to be $3$ for K-means clustering. We show the two-component T-SNE representation~\cite{maaten2008visualizing}, along with the cluster label for each city to show the efficacy of the clustering in Fig~\ref{fig:temporal_diurnal} (b). From the figure, we can see that the clustering so obtained separates the cities well. In the $3$ clusters we obtained, we observed that first cluster corresponded to most European cities (containing $80\%$ of European cities we analyzed). The second cluster consisted only of Indian ($7$) and Middle-Eastern cities ($12$). The final cluster consisted of primarily American cities (containing $86\%$ of American cities we analyzed.

%From the figure, it can be seen that the clustering so obtained separates the cities well. We observe only one outlier cluster, containing only one member i.e., Tripoli. This is because Tripoli has the highest fraction of driving snaps ($0.74$), and hence the temporal patterns exhibited are outliers. The next smallest cluster is the densest cluster, containing mostly popular Asian cities - Lahore, Karachi, Islamabad, Dubai, Jeddah, Mecca, Riyadh, Medina, Doha, Jalandhar, Chandigarh, Ahmedabad, Surat, and New Delhi. All the US cities are clustered together (Cluster $2$ contains all US cities), and similarly most of the European cities belong to the same cluster. (Cluster $1$, containing $81\%$ of the cities we analyzed in Europe). 

%\pk{can we have some quantification for this?}\hl{yes- let's compute NMI by clustering the cities based on different geographical patterns (gt clusters) and then comparing nmi with the clustering we get?}

\begin{observation}[Temporal Clustering]
Temporal patterns exhibited by different cities can be meaningfully clustered, and indicate overall geographical and cultural patterns.
\end{observation}

We can observe the presence of temporal patterns in \driving posting behavior, which answers RQ\ref{rq4}. This analysis could be used by platform designers or policy makers to target cities at a specific time of the day by discouraging or warning users about this type of behavior.

\subsection{Spatial Analysis}
%\hl{Update this section - since grid size has changed - and what is the key insight we are getting?}
Previously, spatial analysis on SnapMaps has been used to show that usage of Snapchat, while posting publicly to maps has been concentrated \cite{juhaszanalyzing}. We use spatial analysis to investigate these insights further while focusing on \driving posting behavior.

\begin{figure}[t]
\centering
    %\hfill
%    \subfigure[]{\includegraphics[width=0.495\linewidth]{images/spatial/Spatial_Fits_NEW.pdf}}
    \subfigure[]{\includegraphics[width=0.495\linewidth]{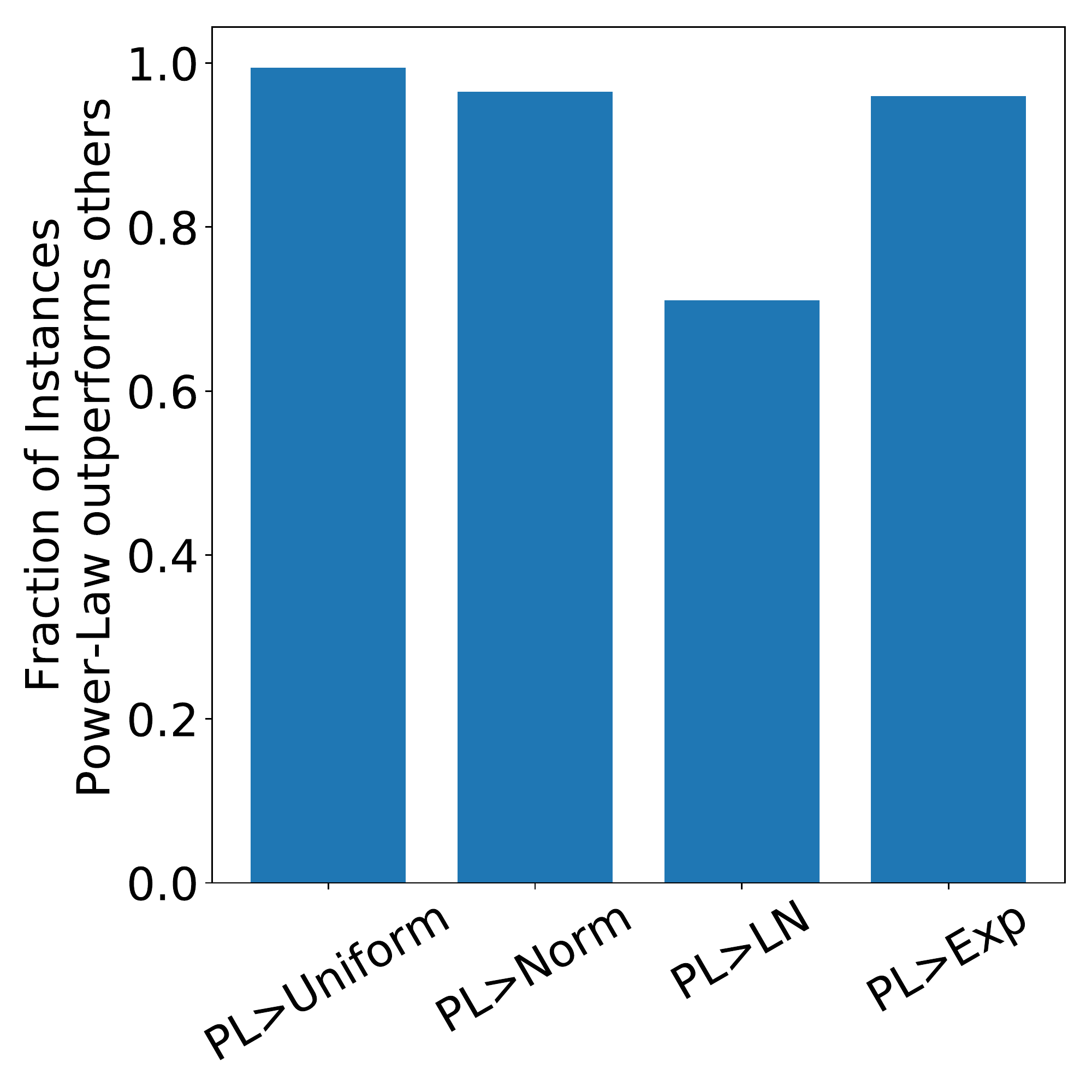}}
    %~\hfill
%    \subfigure[]{\includegraphics[width=0.495\linewidth]{images/spatial/Spatial_Riyadh_Delhi_NEW.pdf}}
    \subfigure[]{\includegraphics[width=0.495\linewidth]{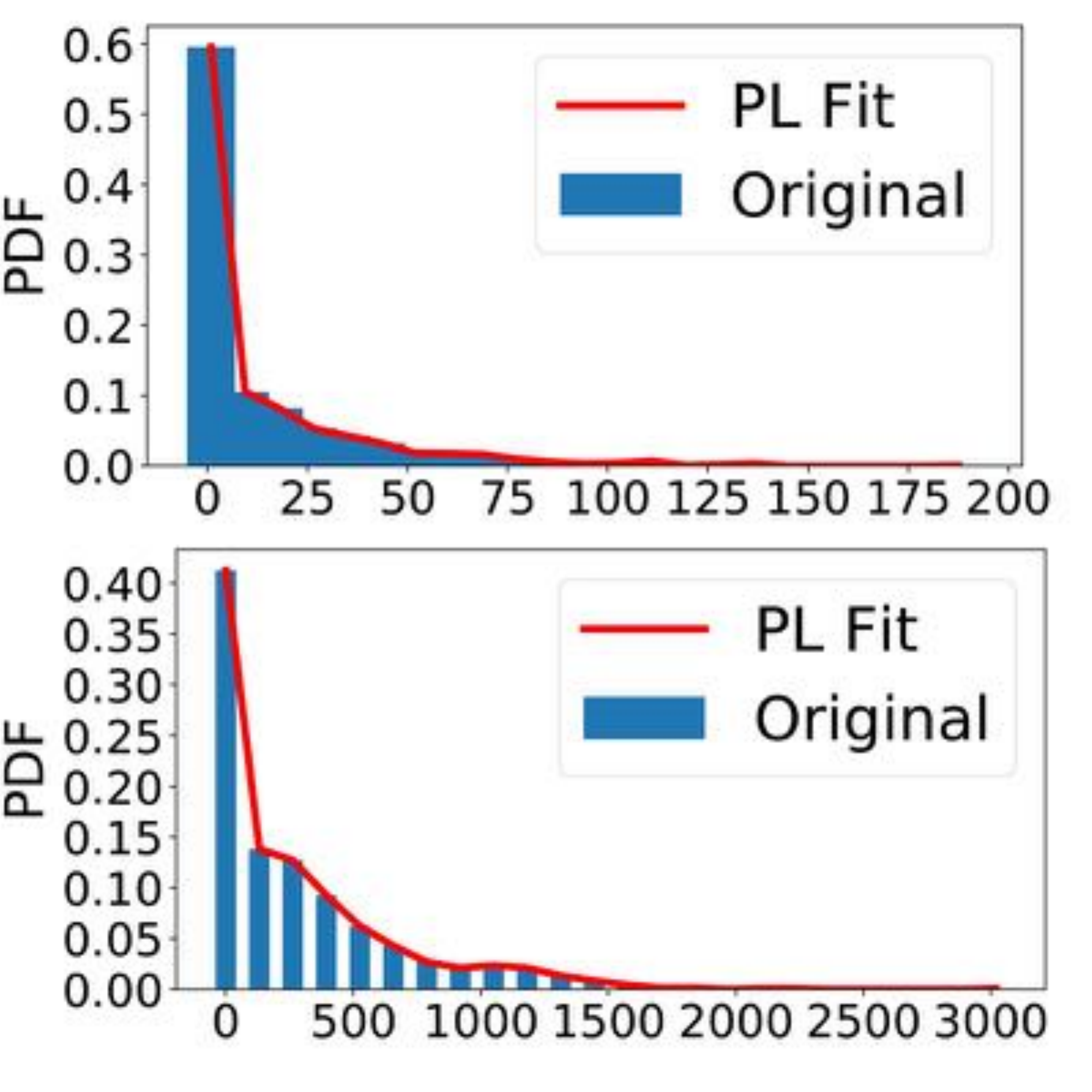}}
    \caption{(a) Power-Law distribution fits the best for most of the cities, in comparison to other candidate distributions. (b) Sample fits under Power-Law distribution shown for (top) Delhi and (bottom) Riyadh.}
    \label{fig:spatial_figs}
\end{figure}

% \begin{figure}[!hbtp]
%     \centering
%     \begin{subfigure}{width=0.225\textwidth}
%         \includegraphics{images/spatial/Spatial_Fits.pdf}
%     \end{subfigure}
%     \begin{subfigure}{width=0.225\textwidth}
%         \includegraphics{images/spatial/Spatial_Riyadh_Delhi.pdf}
%     \end{subfigure}
%     \includegraphics{}
%     \caption{Caption}
%     \label{fig:my_label}
% \end{figure}

% \begin{table}{}
%     \centering
%     \begin{tabular}{cc}
%     \multirow{2}{*}[0.6in]{\includegraphics[scale=0.175]{images/spatial/Spatial_Fits.pdf}} & 
%     \includegraphics[width=0.49\linewidth]{AuthorKit18/LaTeX/images/spatial/Spatial_Delhi.pdf}\\
%          & \includegraphics[width=0.49\linewidth]{AuthorKit18/LaTeX/images/spatial/Spatial_Delhi.pdf}
%     \end{tabular}
    
% \end{table}
    
% \begin{figure}[!b]
%     \centering
%     \includegraphics[scale=0.3]{images/SD_Spatial_Driving.pdf}
%     \caption{Histogram of Standard Deviation (SD) for percentage of driving snaps across tiles for all cities.}
%     \label{fig:concentration}
% \end{figure}

In Figure~\ref{fig:spatial_heatmaps}, we show the spatial distribution of \driving snaps for three popular cities ((a) Delhi, (b) Riyadh and (c) New York City). We can see that for these cities the distribution is concentrated on small regions on the map. To measure if the \driving snaps are concentrated or not, we model the distribution of the number of \driving snaps per tile for each city with a known parametric family of distributions. Concentrated \driving snaps will follow a power-law (PL) distribution, as compared to uniform \driving snaps which will follow a uniform distribution. We try to fit multiple distributions (power-law, gaussian, log-normal, and exponential) on all cities per tile to model \driving content distribution. We discover that power-law distribution fits better than all the other candidate distributions for the majority of the cities when compared using log-likelihood and BIC metrics. We plot the percentage of instances for which power-law distribution fits better than other candidate distributions in Figure~\ref{fig:spatial_figs}(a). We also show power-law distribution fit for two cities - Riyadh (top) and Delhi (bottom) and observe that the fits are visually accurate.

%we compute the variance in the number of driving-related snaps posted across all grids for every city. A high standard deviation indicates that the snaps are concentrated, and a low standard deviation indicates the opposite. We show the standard deviation for the percentage of driving snaps for all the cities as a histogram plotted in Figure~\ref{fig:concentration}. The figure reveals that a large number of cities have higher standard deviations for tiles where driving content was posted. Further investigation reveals that for cities which have at least 50 grids, over half the driving content posts could be traced to just $20\%$ of the tiles.%into the contribution of driving snaps across tiles of the grid of a city reveals that for cities with at least 50 tiles (a data collection area of 312.5 sq. km), over half the driving content posts could be traced to just 20\% of the tiles of the city.

\begin{observation}[Concentrated Driving Content]
For most of the cities across the world, the \driving posting behavior is geographically concentrated to only a few tiles and not uniformly distributed across the city.
\end{observation}

Another interesting pattern that we observe was that for certain cities, the \driving was observed to be higher on major roads. For example, in Riyadh's heatmap (Figure~\ref{fig:spatial_heatmaps}(b)), we can see two major roads having a higher concentration of distracted driving snaps. However, we cannot quantify this pattern across all cities as we do not have access to underlying road and highway data and leave this pattern quantification as future work.

We discovered useful insights about distracted driving content posting behavior within and across cities, thus answering RQ\ref{rq3}. Such insights can be used to develop interventions based on geographic areas.

\section{Characterizing Users}
For our investigation into the demographics of the user (RQ\ref{rq2}), we aim to understand how the demographics of a particular city affect the number of driving snaps.

\subsection{Explanatory Variables}
Most previous work in risk-taking has focused on two important characteristics of individuals indulging in risk-taking activities~\cite{leary1994self,edgework1990lyng}, namely gender and age. In this work, we extend their work and investigate the role of gender and age in a user's proclivity to create distracted driving content. Therefore, we examine these two features - gender and age distribution for each city. Additionally, since Snapchat is a popular Internet-based platform, it is imperative to understand the economic influences that might affect the type of usage of the platform. Therefore, we use the development status of a country in which the city is as one of the control variables. We classify the countries of the world in our dataset as either \emph{developed} or \emph{developing} based on the definition of developed nations given in CIA's world factbook~\cite{central2009cia}. The economic status of the city further acts as a proxy for various other additional variables for which data is less readily available such as smartphone penetration, social media usage, and availability of public transportation facilities. We also account for certain control variables such as the total number of snaps posted in the city and the population of the city. We obtain the population estimate for each city from \url{worldpopulationreview.com}, where we use the latest estimate available. Similarly, we obtain the gender ratio statistic from the latest available census data that has been aggregated on \url{citypopulation.de}. However, the website does not provide us with the latest data for all the cities. In such cases, we take the latest gender ratio available and assume that it remains constant for the city. For computing the age-distribution, we used the statistics from \url{citypopulation.de}, and for cities where the data was not available - census data for the respective country was obtained. It is possible that for statistics such as gender and age, the statistics across cities might have been computed for different years. To account for this discrepancy, we use age and gender variables as a percentage over the total population. Finally, we did not include cities for which we did not have satisfactory census data, which left us with $130$ cities.
%Finally, we did not include the cities for which we did not have satisfactory data, which left us with $101$ cities for which all of these variables were available. 
%\hl{Mention how obtaining this data was a challenging task - and we will make it public}`

\subsection{Effect of Variables}
\begin{figure}
\centering
    %\hfill
%    \subfigure{\includegraphics[width=0.23\textwidth]{images/regression/male_ratio-converted.pdf}}
    \subfigure{\includegraphics[width=0.23\textwidth]{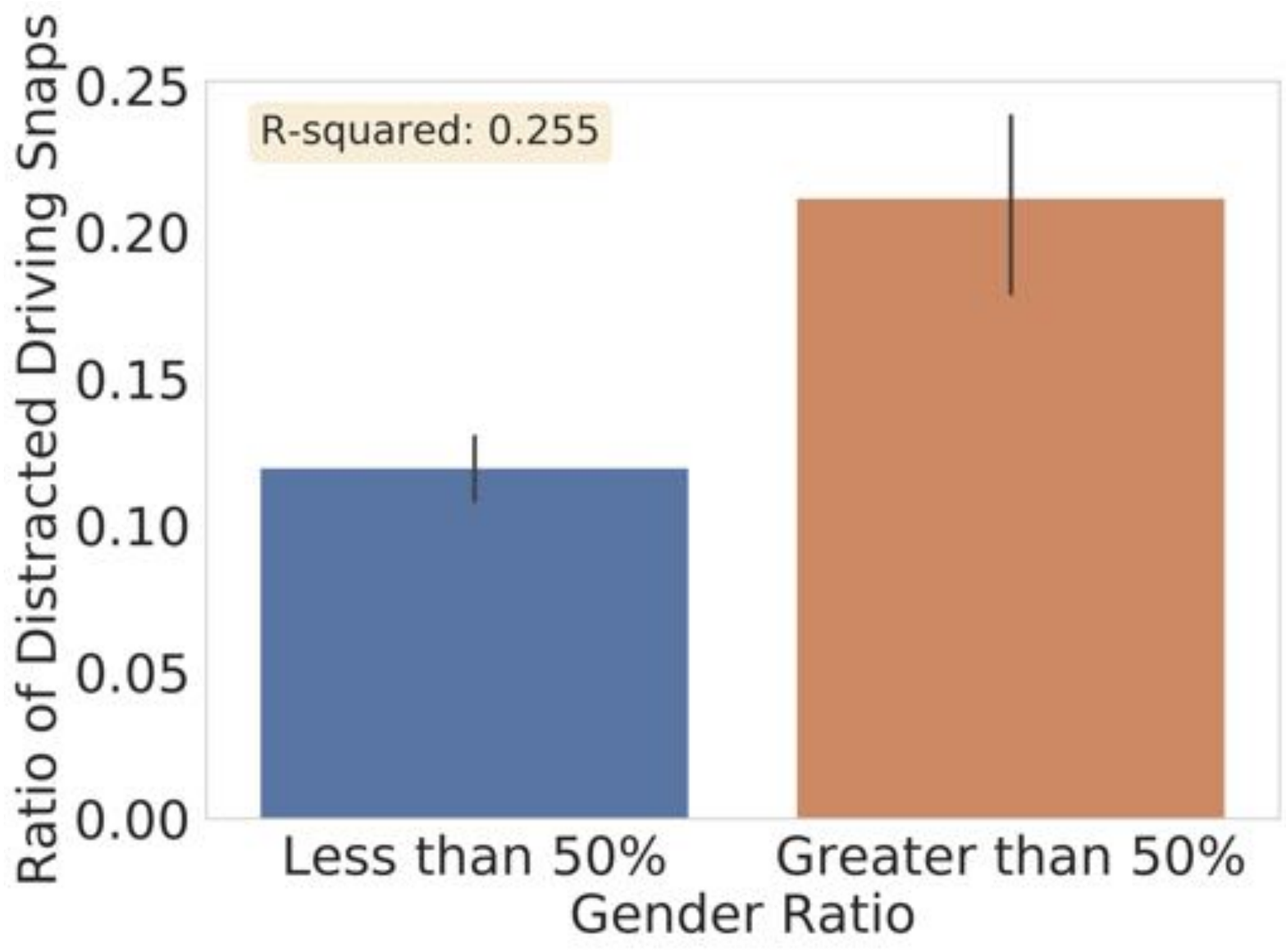}}
%    \subfigure{\includegraphics[width=0.23\textwidth]{images/regression/developed_vs_developing-converted.pdf}}
    \subfigure{\includegraphics[width=0.23\textwidth]{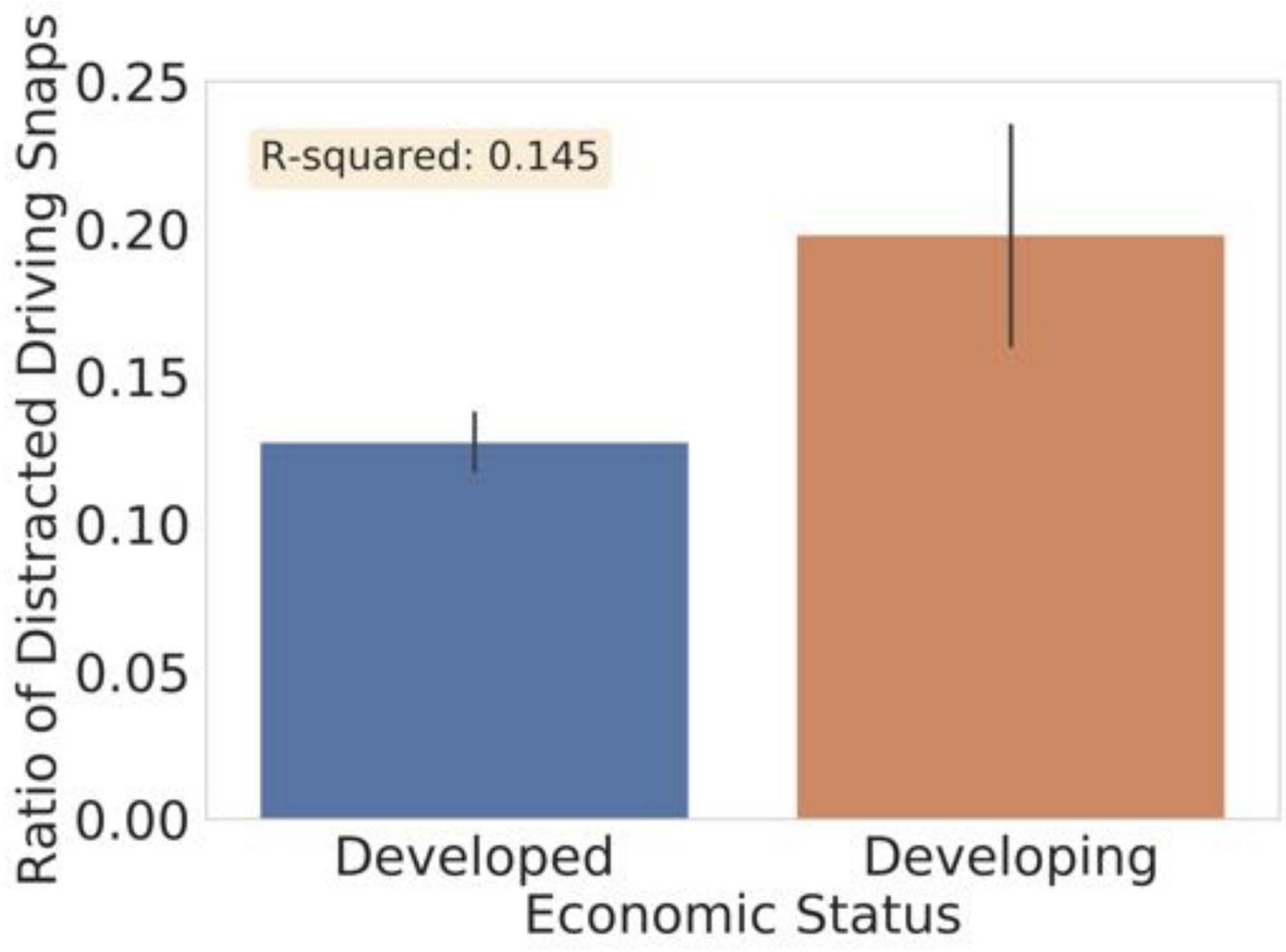}}
    ~\hfill
%    \subfigure{\includegraphics[width=0.23\textwidth]{images/regression/population_log-converted.pdf}}
    \subfigure{\includegraphics[width=0.23\textwidth]{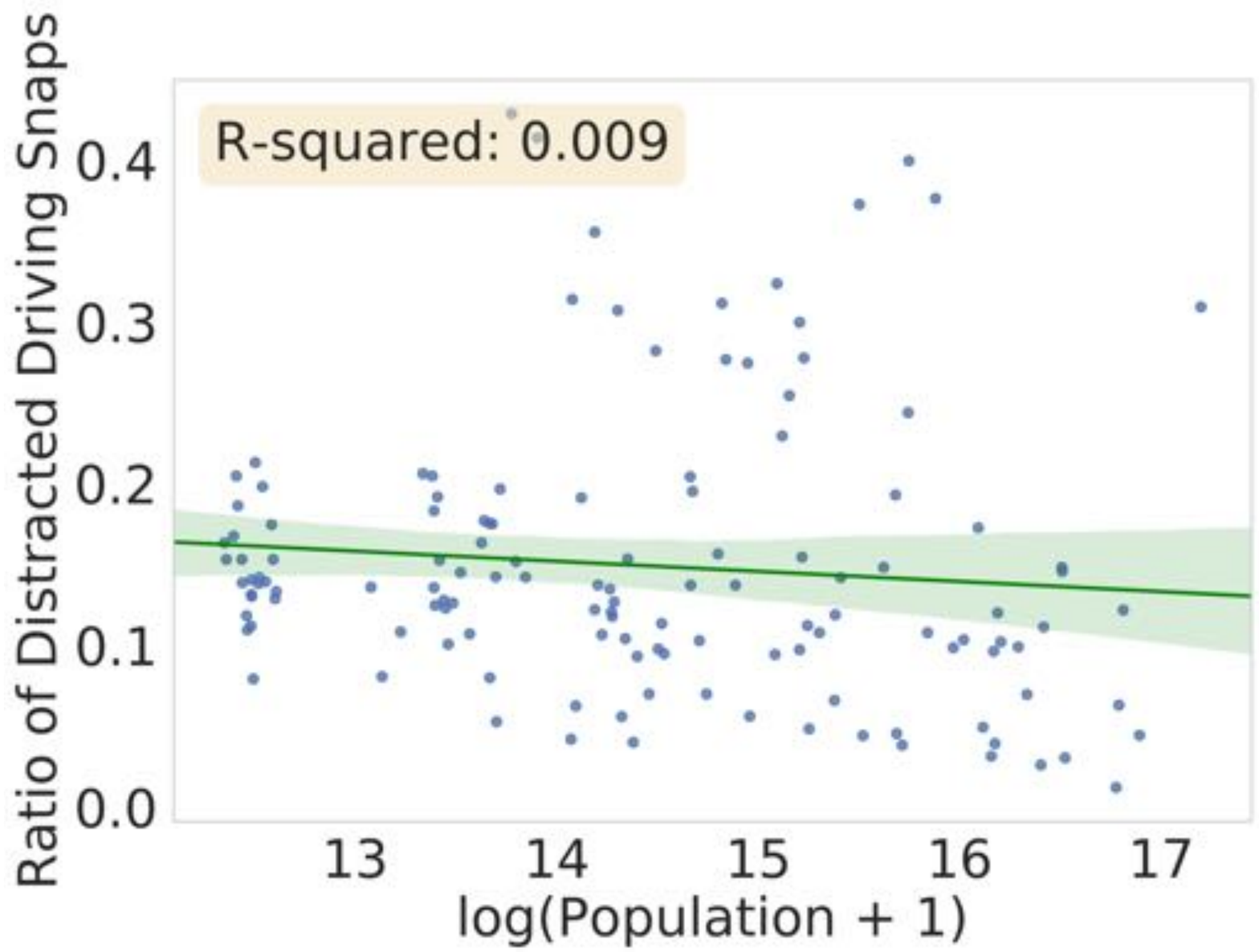}}
%      \subfigure{\includegraphics[width=0.23\textwidth]{images/regression/age_log-converted.pdf}}
      \subfigure{\includegraphics[width=0.23\textwidth]{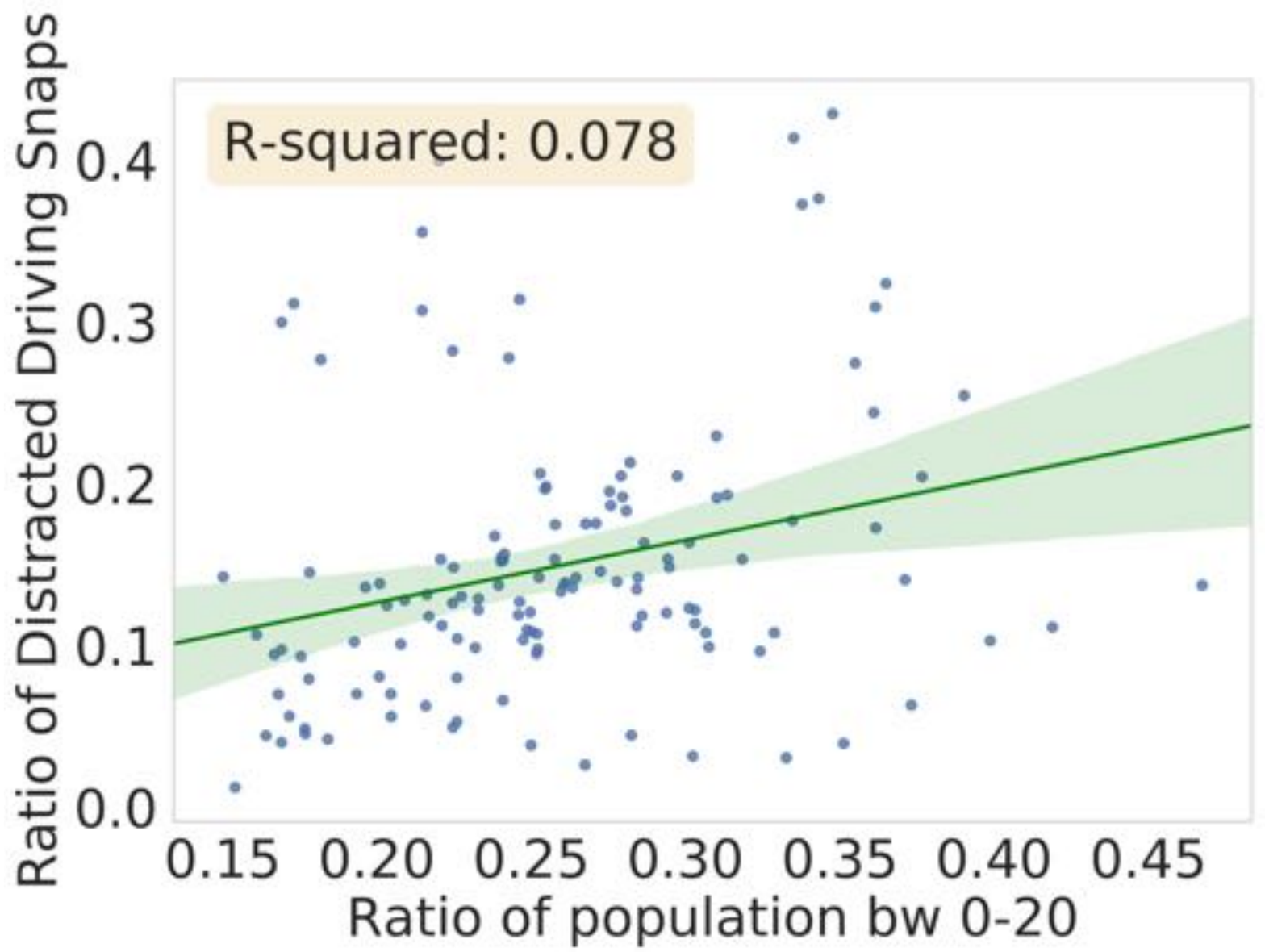}}
    \caption{Scatter plot of how number of driving snaps is affected by different variables: (a) Gender Ratio: Ratio of Males to Females (b) Development status of the city (c) Population of the city, (d) Ratio of population between ages 0 and 20}
    \label{fig:regression_effect}
\end{figure}

We investigate the relationship between the variables mentioned above and the number of distracted driving snaps posted from each city, based on which we observe some interesting patterns. From Figure ~\ref{fig:regression_effect}(a), we can observe that the distracted driving snaps ratio for cities where the gender ratio is in favor of males is roughly $77\%$ more than that of the cities where the gender ratio is in favor of females ($t=6.62$, $p<0.001)$. Similarly, from Figure ~\ref{fig:regression_effect}(b), we observe that distracted driving snaps ratio posted in the developing cities is roughly $55\%$ more than that of the developed cities ($t=4.66$, $p<0.001$). In Figure ~\ref{fig:regression_effect}(c), we present the scatter plot of the population of a city (log scale) with the number of driving snaps posted. We can see that there is a small negative slope, possibly implying that cities with the larger population have a lower number of driving snaps. Interestingly, we note that the slope in the case of the ratio of the population below $0-20$ is positive ($R^{2}=0.078$, $p<0.01$), suggesting that cities with a higher ratio of population in the age group of $0-20$ have a higher number of driving snaps. 

\subsection{Statistical Model}
We are interested in explaining the number of distracted driving snaps posted from every city. We assume a linear relationship between the number of distracted driving snaps and the other variables discussed previously. We transform all the count variables to log-scale to stabilize their variances. The explaining variables (or independent variables) along with the dependent variable we use to model are shown in Table~\ref{table:variables}. Besides the explaining variables - we also use the number of total snaps as a natural control for the popularity of Snapchat in the city. We present the results of the regression on all the $130$ cities for which we were able to get satisfactory data in Table~\ref{table:reg_results}.

\begin{table}[!ht]
    \centering
    \caption{List of dependent variables used to estimate the number of driving snaps posted.}
    \begin{tabular}{llcc}
    \toprule
    Variable Name & Description & Min. & Max.   \\  
    \midrule
    \midrule
    \multicolumn{4}{l}{\textbf{Independent Variables}}  \\
    \midrule
    $\log (Pop. + 1)$ & Population & $12.35$ & $17.19$    \\
    $ Age < 20$ & \% of pop. \textless 20 & $15.0$ & $46.7$    \\  
    $ 20 < Age < 40$ & 20 \textgreater \% of pop. \textless 40 & $19.4$ & $58.3$ \\
    $ 40 < Age < 60$ & \% of pop. \textgreater 40 & $14.1$ & $60.5$  \\
    Male ratio & Ratio of Male pop. & $0.458$ & $0.756$ \\
    $\log (TS + 1)$ & \# Total Snaps & $5.412$ & $13.813$ \\    
    % Grid Size & Size of the city & &    \\
    \midrule
    \multicolumn{4}{l}{\textbf{Dependent Variable}} \\
    \midrule
    $\log (DS + 1)$ & \# Driving Snaps & $2.08$ & $12.89$   \\
    \bottomrule
    \end{tabular}
    \label{table:variables}
\end{table}

\begin{table}[!hb]
    \centering
    \caption{Regression models for number of distracted driving snaps (N=130).}
    \begin{tabular}{lcc}
    \toprule
    \toprule
        & \multicolumn{2}{c}{\textbf{Dependent variable}}    \\
        & \multicolumn{2}{c}{$\log(DS+1)$}  \\
        \cline{2-3}
         & Coeffs(Err.) & LR ChiSq \\
        \midrule
        Intercept & $-6.86 (1.48)^{***}$ & \\
        $Males$ & $0.05 (0.01)^{***}$ & $607.14^{***}$  \\
        $Age<20$ & $5.85 (1.46)^{***}$ & $33.72^{***}$\\
        $20<Age<40$ & $1.92 (1.56)$ &  $0.60$ \\
        $40<Age<60$ & $2.38 (1.40)^{.}$ & $2.89^{.}$ \\
        $Developing$ & $0.19 (0.12)^{.}$ & $81.48^{***}$ \\
        $\log(Pop.+1)$ & $-0.21 (0.03)^{***}$ & $51.53^{***}$   \\
        $\log(TS+1)$ & $1.21 (0.03)^{***}$ & $2269.99^{***}$ \\
        \midrule
        $R^{2}$ coefficient & \multicolumn{2}{c}{$0.9593$} \\
    \bottomrule
    \bottomrule
    \multicolumn{3}{c}{{Note:$^{***}p<0.001$, $^{**}p<0.01$, $^{.}p<0.1$}}
    \end{tabular}
    \label{table:reg_results}
\end{table}

Analyzing the results, we can see that the term Total Snaps (TS) introduced as a control variable behaves as expected. The effect of the variable is significant and positively related, with a one percent rise in the log number of snaps posted associated with a $1.21\%$ rise in the log number of distracted driving snaps. We can also see that the population of a city has a significant negative effect. This could perhaps be explained by the fact that as the cities grow in population, the traffic and congestion on the road also increases, leading to more time spent on paying attention to the road as compared to that spent on a phone.
% thus not allowing users to drive fast and show off their risk-taking personality. 

Connecting back to our RQ\ref{rq2}, we want to figure out what demographics of users are more inclined to indulge in \driving posting behavior. We first investigate the role of gender and its contribution to the number of distracted driving snaps across cities. It has often been shown that proclivity of taking risk is higher among males~\cite{leary1994self,edgework1990lyng}. We verify the same hypothesis in our regression model, where we observe that the percentage of the male population has a significant, positive, and large effect. A one percent increase in the male ratio would lead to a 0.05\% rise in the log number of distracted driving snaps.

\begin{observation}[Role of Gender]
Cities with higher male ratio are more likely to produce more distracted driving snaps.
\end{observation}

Another popular result of the edgework framework is that younger people are more likely to participate and indulge in risk-taking activities. In our model, we introduced $3$ variables as percentage of individuals less than 20 years of age ($Age<20$), between 20 and 40 years of age ($20<Age<40$), and above 40 ($Age>40$). We discovered that $20<Age$ has a significant positive effect on the number of distracted driving snaps posted in the city. However, the other two variables did not have any significant effect. Though this result is significant, it is also probably biased as Snapchat is a platform that is primarily used by young people; hence, there is a possibility that this observation just might be capturing that effect.

%However, we discovered that this variable had no significant effect on the number of driving snaps, thereby refuting the previous hypothesis for the case of driving snaps over these cities. We believe that this could be because most of the youngsters do not get access to their own vehicles, especially in developing cities, until they have graduated from college. It further can be hypothesized that once an individual reaches the age of 25, and probably has access to their own vehicle, the individual already is out of the fascination towards risk-taking endeavors and is more careful.
\begin{observation}[Role of Age]
Cities with higher proportion of young people are more likely to post distracted driving snaps than cities with higher proportion of older people.
\end{observation}

Additionally, we see that there is an effect of whether the city is developed or developing ($Developing$) on the number of distracted driving snaps that get posted. We discover that if a city is in a developing nation, then there are higher chances of distracted driving snap posting behavior. This is in accordance with the overall spatial and temporal pattern observed, the cities being ranked consistently higher in distracted driving snap posting behavior were mostly cities from developing countries. 

%Similarly, we see a negative effect of the mean income. This implies that more the mean income of the city, lesser the number of driving snaps will be posted from that city. We also experimented with developed/developing nation, and found it to be highly correlated with mean income, hence omitted that from the model. We are unaware of the exact reason behind this phenomenon, and would require more careful analysis.
\begin{observation}[Effect of Development]
Users from cities in developing world are more likely to post distracted driving snaps.
\end{observation}

%In previous works of risk-taking, it has often been shown that proclivity of taking risk is higher among males. We verify the same hypothesis in our regression model, where we observed that the percentage of male population has a significant, positive and large effect. A one percent increase in the male population would lead to a 2.2\% rise in the number of driving snaps.
\section{Discussion}

\subsection{Research Questions}
\textbf{RQ}\ref{rq1} relates to the extent of distracted driving snaps are posted on Snapchat across cities. The question tries to estimate the prevalence of such type of risk-taking behavior on social media platforms, thus quantifying the importance of studying such problems. We discovered that distracted driving snaps form $23.56\%$ of total snaps posted across $173$ cities. Further, we also noticed that such behavior is more prevalent in Middle-Eastern and sub-continent Indian cities(accounting for $72.4$\% of distracted driving snaps overall). By answering \textbf{RQ}\ref{rq3}, we investigated the spatial patterns of distracted driving content posting behavior. We discovered that such content is posted in certain regions of the city; and is not uniform across the city, thus, showing that distracted driving content posting behavior is concentrated. However, we were unable to analyze these hotspots for the underlying demographic and geographical features to understand the reason behind such concentration - largely due to the lack of data at that granularity. \textbf{RQ}\ref{rq4} is focused on determining temporal patterns behind distracted driving content posting behavior. We made key observations based on temporal analysis of the behavior across cities. We discovered that most of such content is posted heavily during night-time. Further, we were also able to discover strong regional effects - where the clusters formed on clustering the fraction of snaps posted each hour of the week segmented into clusters comprising majorly of European, American and Mid-Eastern cities.

One of the key frameworks proposed by sociologists to explain risk-taking literature has been \emph{edgework}. The framework, besides defining voluntary risk-taking behavior and applying it to different settings, also proposed characteristics that define the users who are inclined to take such risks. The observations made about such voluntary risk-takers was based on the concept of an illusory sense of control, where a user feels that they have more control of the situation than they actually do. The theory discovered that males and young people generally felt more of such an illusory sense of control. We tested whether the theories put forward by the \emph{edgework} framework also hold for the case of distracted driving content posting behavior on social media platforms. We attempted to answer this in \textbf{RQ}\ref{rq2}. We discovered, in concurrence with the theory, that males are more inclined to participate in such voluntary risk-taking behavior. Further, we also discovered that younger people are more inclined to exhibit such behavior, another key characteristic proposed by the framework. Another key point put forth by the theory was that individuals who were of a social system that exhibited much larger control over their life ended up participating in such behavior in seek of a high-stakes feeling of control over the situation. We hypothesized that this could relate to the economic situation of a particular city - and tested if individuals from developing regions (instead of developed) were more likely to participate in risks or not. We discovered that we do see the effect of the economic status of the city. However, we only treat economic status as a proxy for control; many other factors such as political and cultural could be considered, which are hard to obtain and quantify.

Finally, to be able to answer any of the \textbf{RQ}s as mentioned earlier, we needed to figure out how can we detect if a particular snap is an example of distracted driving content or not. Due to the large scale of our study, it is infeasible to label the entire dataset manually. Hence, we answered \textbf{RQ}\ref{rq0} by proposing a deep learning classifier and were able to achieve high precision and recall. Further, we even tested the robustness of the trained classifier to show that the proposed method performs robustly on an held-out set.

%Previous work in risk-taking literature has been done from a sociological perspective. The major framework proposed 
%We discovered interesting spatial patterns - finding that the content posted across cities varies drastically - more specifically distracted driving content is concentrated on small areas in the city; rather than being uniformly distributed. For temporal analysis, we discovered that distracted driving snap posting behavior was diurnal and the overall temporal patterns are clustered together by their geographical perimeter of the city. Lastly, we investigated \textbf{RQ}\ref{rq2}, where we tried to answer how does demographics of a city influenced distracted driving content posting.
\subsection{Implications}
Our paper provides a robust way of detecting if the content posted on Snapchat is an instance of distracted driving content or not. Further, our results provide insights into the extent of such behavior on a popular social media platform Snapchat, and spatial, temporal and demographics related patterns. We believe that the platform owners and policymakers can leverage insights put forward by our work to develop educational campaigns and interventions. We discuss some of the suggestions below:

\noindent
\textbf{Location-Based}: One of our key insights (Insight 1) was that distracted driving content posting behavior is prevalent mostly in Middle Eastern and Indian cities. Thus, some of the educational campaigns could be focused only on these regions and can be disseminated within the platform itself. Another insight that could be crucial in designing platform-based interventions is that such behavior is concentrated only in certain regions of cities. The platform owners can analyze the content posted around these hotspots with the proposed deep learning classifier to determine if the content posted are instances of distracted driving or not, and make a decision of not showing such content at all. In the case of Snapchat specifically, users post such content on SnapMaps to gain popularity from the general public; however, if such content is not allowed to be posted on the platform from these regions, there is a possibility that it might discourage the individuals from creating such content. However, this requires more experimentation to determine if such a form of intervention can be useful or not.

\noindent
\textbf{Time-Based}: Our work made a useful insight about nighttime driving, indicating that such content is generally posted late in the night (insight 2). This insight could be leveraged to issue educational notifications at that time of the day when such at-risk users could be active.

\noindent
\textbf{Demographics-Based}: The major insight we draw from our regression analysis was the role of age and gender in characterizing the users who participate in such behavior. We discovered through insights 5 and 6 that young individuals and males are more likely to participate in such behavior. If a platform has a way of inferring identities of their users, it could be leveraged in combination with the other insights to create targeted interventions and educational campaigns for these specific demographics. 

We are aware that a social media platform has other constraints while issuing notifications, such as the number of them, and restricting users not to share certain types of content, which could potentially lead to violation of their freedom to express. All the above mentioned interventions can be combined with the proposed deep learning classifier to give the platform owners more flexibility to design interventions and educational campaigns. Since such interventions can also act in unintended ways (such as suggesting risk-takers not to perform risk-taking behavior; hence, actually motivating them), more analysis needs to be done before proceeding forward.

\subsection{Threats to Validity}
Like any quantitative study, our work is subject to threats to validity. We try to enumerate biases, issues, and threats to the validity of our study by following a framework for inferring biases and pitfalls while analyzing social data by Olteanu et al.\cite{olteanu2019social}. First, our work is based on the data collected on Snapchat, mostly through SnapMap. A key data issue is that of representativeness - our collected data, though might not be geographically or temporally biased (since we collected data across the world and for a large amount of time), it can still be that we are collecting data disproportionately from regions that post more frequently publicly on SnapMaps rather than Snapchat in general. Another representative issue is that we are linking Snapchat usage data with that of census data in general; where Snapchat users might not be representative of the entire cities population. We try to discount this representation bias by including appropriate control variables, but still, some of the bias might exist in our analysis. Additionally, our dataset might also contain temporal bias as during our one-month long data collection; it might be possible that some cities might be observing festival-related holidays or some events. This might have introduced a disproportion in the number of snaps collected from each city. A significant source of data bias in our analysis is the use of census data. Firstly, we were not able to obtain data for each city and thus had to omit certain cities from our analysis. Secondly, census data is obtained from different years, and finally, the census data for different cities are taken from different sources.

For the annotation required for training deep learning classifier, we used a limited number of annotators, which might result in subjective interpretation. We attempted to mitigate this threat by using majority voting and computing inter-annotator agreement rate. Finally, our statistical modeling required multiple parameters that were related to the operationalization of theories that exist in literature. Some of these parameters might not be capturing the factors that we intended to capture or that the theories captured. Additionally, we made an assumption where we posited content posted by a front seat passenger also as a form of distracted driving content, which might not be true. It could be possible that our analysis might be applicable only for Snapchat and might not generalize well for other platforms and also for other risk-taking behavior.

\section{Related Work}
Besides the relevant theories and framing discussed in ``Development of Research Questions", there are other related work that should be discussed. We discuss them here:

\noindent
Recently, there have been some studies on analyzing risk-taking behavior on social media for different voluntary activities. Lamba et al. covered a much broader case of dangerous selfies, where users posted a perilous self-portrait in dangerous situations such as at an elevation, with a firearm, or inside a water body\cite{lamba2017camera}. They also showed that users often engage in risk-taking activities while taking selfies to post on social media. Of the 232 deaths due to taking dangerous selfies, $12$ could be attributed to driving-related incidents. The authors presented deep-learning models to distinguish between potentially dangerous and non-dangerous selfies~\cite{nanda2018stop}. Similarly, Hart examined young individuals' participation in posting nude self-portraits on Tumblr \cite{hart2017being}. There has been a normative increase in individuals dabbling in risk-taking behavior as a result of various other social media trends such as the Tide Pod Challenge \cite{murphy2018rationality}, the Cinnamon Challenge \cite{grant2013ingesting}, the Salt and Ice Challenge \cite{roussel2016tweens} and the Fire Challenge \cite{ahern2015risky,avery2016fire}. However our work is the first in analyzing the specific behavior of distracted driving content posting on social media. Further we extend the popular voluntary risk-taking \emph{edgework} framework to social media platforms.

\section{Future Work}
In this work, we concentrated on characterizing the extent of distracted driving content posted on Snapchat. However, this study could be extended to other platforms as well. Additionally, we concentrated on corroborating edgework framework for distracted driving content - this could potentially be extended to other edgework activities that can be observed on online social platforms. Technically, we made the assumption that front seat passenger posted video can potentially be also dangerous - however the classifier can be made only to annotate driver posted content as true positive by making the architecture either more hierarchical (first classify whether the content is about driving and then if it is posted by driver or not) or by carefully annotating the training set.

\section{Conclusions}
In this work, we investigate the widespread prevalence of distracted driving content posting behavior. We specifically focus on a popular social media platform, Snapchat, and by analyzing the publicly posted stories, we characterized the extent of distracted driving content that exists on such platforms.

Our first contribution is proposing a deep learning based classifier to detect if a content posted is distracted driving or not. Grounding our work in risk-taking literature, we aim to test out the theories put forth by sociologists in terms of risk-taking behavior in the offline world in the context of distracted driving content posting behavior on social media platforms and test them. To this end, we proposed and answered multiple RQs related to extent, spatial, temporal and demographic patterns of such behavior across $173$ cities.

We made the following key observations related to the few RQs - the demographics such as age and gender play a key role in the proclivity to post distracted driving content. Further, we also discovered that there exists spatial and temporal patterns in distracted driving content behavior posting across cities. We hypothesize that the insights derived from this study can be used to design targeted intervention and educational campaigns to curb such risk-taking behavior.

\vspace{1em}
\noindent
\footnotesize{\textbf{Acknowledgements}
We are very thankful to Bogdan Vasilescu for his comments on the draft. We are also very thankful to the anonymous reviewers for their thoughtful suggestions and comments.
}

% References and End of Paper
% These lines must be placed at the end of your paper
\bibliography{references}
\bibliographystyle{aaai}

%\input{response}
%\newpage
%\newpage
%\input{supplementary}

\end{document}